\documentclass[epsf,twocolumn,10pt]{IEEEtran}
\usepackage{multirow}
\usepackage{amssymb}
\usepackage[dvips]{graphicx}
\usepackage[cmex10]{amsmath}
\usepackage{amsthm}
\usepackage{latexsym,bm}
\usepackage{color}
\usepackage{subfigure}
\usepackage{lettrine}
\newcommand{\E}{{\rm E}}

\theoremstyle{plain}

\theoremstyle{definition}
\theoremstyle{definition}

\usepackage{epstopdf}
\usepackage{graphicx}

\begin{document}

\title{Design of an MISO-SWIPT-Aided Code-Index Modulated Multi-Carrier $M$-DCSK System for
e-Health IoT
\thanks{Manuscript received December 1, 2019; revised February 25, 2020; accepted March 15, 2020. Date of publication
Month XX, 2020; date of current version Month XX, 2020. This work was supported in part by the NSF of China (Nos. 61701121, 61771149, 61871132, 61871136, 61801128), the Open Research Fund of National Mobile Communications Research Laboratory, Southeast University, under Grant 2018D02, the NSF of Guangdong Province (No. 2019A1515011465), the Science and Technology Program of Guangzhou (No. 201904010124), the Guangdong Province Universities and Colleges Pearl River Scholar Funded Scheme (No. 2017-ZJ022), the Research Project of the Education Department of Guangdong Province (Nos. 2017KTSCX060, 2018KTSCX057, 2017KZDXM028), the Graduate Education \& Innovation Project of Guangdong Province (No. 2020SQXX12), the Guangdong Innovative Research Team Program (No. 2014ZT05G157), the Project of Department of Education of Guangdong Province (No. 2018SFJD-01). {\em (Corresponding author: Yi Fang.)}}
\thanks{G. Cai and G. Han are with the School of Information Engineering, Guangdong University of Technology, Guangzhou 510006, China (e-mail: \{caiguofa2006, gjhan\}@gdut.edu.cn).}
\thanks{Y.~Fang is with the School of Information Engineering, Guangdong University
of Technology, Guangzhou 510006, China, and also with the National
Mobile Communications Research Laboratory, Southeast University, Nanjing
210096, China~(e-mail: fangyi@gdut.edu.cn).}
\thanks{P.~Chen is with the Department of Electronic Information, Fuzhou University, Fuzhou 350116, China (e-mail: ppchen.xm@gmail.com).}
\thanks{G. Cai is with Department of Neurology, Fujian Medical University Union Hospital, Fuzhou 350001, China (e-mail: cgessmu@fjmu.edu.cn).}
\thanks{Y. Song is with the School of Electrical and Electronic Engineering, Nanyang Technological University, Singapore 639798; and also with the School of Information Engineering, Guangdong University of Technology, Guangzhou 510006, China (e-mail:songy@ntu.edu.sg).}
}

\author{Guofa Cai, {\em Member, IEEE}, Yi Fang, {\em Member, IEEE}, Pingping Chen, {\em Member, IEEE}, \\ Guojun Han, {\em Senior Member, IEEE}, Guoen Cai, and Yang Song}

\maketitle

\begin{abstract}
Code index modulated multi-carrier $M$-ary differential chaos shift keying (CIM-MC-$M$-DCSK) system
not only inherits low-power and low-complexity advantages of the conventional DCSK system, but also significantly increases the transmission rate. This feature is of particular importance to Internet of Things (IoT) with trillions of low-cost devices.
In particular, for e-health IoT applications, an efficient transmission scheme is designed to solve the challenge of the limited battery capacity for numerous user equipments served by one base station.
In this paper, a new multiple-input-single-output simultaneous wireless information and power transfer (MISO-SWIPT) scheme for CIM-MC-$M$-DCSK system is proposed by utilizing orthogonal characteristic of chaotic signals with different initial values.
The proposed system adopts power splitting mode, which is very promising for simultaneously providing energy and transmitting information of the user equipments without any external power supply.
In particular, the new system can achieve desirable anti-multipath-fading capability without using channel estimator.
Moreover, the analytical bit-error-rate expression of the proposed system is derived over multipath Rayleigh fading channels.
Furthermore, the spectral efficiency and energy efficiency of the proposed system are analyzed.
Simulation results not only validate the analytical expressions, but also demonstrate the superiority of the proposed system.

\end{abstract}


\section{Introduction}
\IEEEPARstart{E}-{HEALTH} Internet of Things (IoT) is considered as one of the most important missions for 5G network \cite{1}.
With the continuous growth of the e-health IoT, a number of low-cost and low-power devices are connected to the 5G network for well adapting to the increased application domains and deployments \cite{2}. One fundamental aspect of e-health IoT communications is to perform the network in a self-sufficient way, thus rendering the battery lives of the medical devices up to ten years \cite{3}.
Recently, wireless information and power transfer (SWIPT) system, which can simultaneously carry information and energy, has attracted significant attention as it can provide energy supply for the power-limited devices by using radio frequency (RF) signals.
Hence, SWIPT system is considered as a promising candidate for the energy-constraint e-health IoT.

In the SWIPT system, two practically realisable receiver structures, i.e., time switching (TS) and power splitting (PS), have been proposed \cite{4}. In the former, the receiver switches over time between energy harvesting and information decoding, however, in the latter, the receiver uses a part of the received energy for energy harvesting and the remaining one for information decoding.
In the past decade, a large volume of research works have been extensively studied to designing SWIPT schemes with these two receivers in a variety of scenarios \cite{4-1}, e.g., single-input single-output (SISO) system \cite{5,6}, multiple-input multiple-output (MIMO) system \cite{4,7,8}, multiple-input single-output (MISO) multi-user system \cite{9}, cognitive radio networks \cite{10,11,12}, multihop networks \cite{13,14,15}, multi-relay systems \cite{16,17}, and the practical modulation system \cite{18}-\cite{22}. 
However, the above coherent SWIPT systems require the availability of channel state information (CSI), which results in an increase of implementation complexity and the energy consumption due to channel estimation.

In contrast, the noncoherent SWIPT systems, which require no channel estimation, have appeared to be a lower-complexity alternative to the coherent SWIPT systems \cite{23}-\cite{26}. Hence, they have been considered as a competitive solution to e-health IoT applications. 
Among various non-coherent modulation techniques, for example, frequency-shift-keying (FSK) \cite{23}, differential phase shift keying (DPSK) \cite{24}, and differential chaos shift keying (DCSK) \cite{25,16}, DCSK has drawn increasing attention due to its stronger anti-multipath fading capability.\footnote{In fact, IoT communications can be separated into two categories: short-distance wireless communications and low-power wide-area communications. DCSK modulation is a promising physical-layer technique for short-distance wireless communications.}
Moreover, the chaotic signal used in the DCSK system outperforms the conventional single-carrier signal in terms of the RF-to-direct-current (RF-DC) conversion efficiency at the same average transmitted power in the wireless power transfer system \cite{25-1}.
Nevertheless, the main disadvantage of the DCSK system is that a half of the symbol energy is consumed in the reference chaotic signal, which leads to relatively low energy efficiency and transmission rate.
To overcome this problem, many DCSK variants have been proposed to improve the transmission rate and energy efficiency \cite{27}-\cite{33}, and some coded DCSK schemes have been conceived in \cite{34,35} by using some modern codes, e.g., channel coding \cite{36} and network coding \cite{37}, to improve the transmission reliability, thus facilitating its applications in various communication scenarios.

Recently, the joint design of the DCSK modulation and the index modulation have been proposed to further increase the transmission rate and energy efficiency.
To inherit the noncoherent property of the DCSK system, a differentially spatial modulated chaos shift keying modulation has been conceived by using a differential matrix \cite{38}, where the antenna index is employed to carry additional information bits.
In \cite{39}, a multi-user permutation-index DCSK system has been proposed by permutation-matrix to achieve high transmission rate and enhance system security.
In \cite{40,41,42}, the code-index-modulation (CIM) DCSK systems have been designed through selecting different Walsh codes to increase the transmission rate and to improve energy efficiency.
Furthermore, with respect to the single-carrier communication systems, multi-carrier (MC) communication system has been employed in the current communications, which is also a promising proposal for 5G network, especially, in IoT applications \cite{43}.
Based on MC-DCSK system \cite{30}, a carrier-index MC-DCSK (CI-MC-DCSK) system and its $M$-ary version have been conceived via the subcarrier index to further reduce the energy consumption \cite{44,45}. In particular, compared with the PI-DCSK and CIM-DCSK systems, the CI-MC-DCSK system can significantly reduce the system complexity due to the RF delay lines.
Following the CI-MC-DCSK system, a new CIM-MC-$M$-DCSK system utilizes Walsh codes in the reference chaotic signals for all subcarriers to convey additional information bits and ensure the transmission information for all subcarriers \cite{46}.
For the above reasons, the CIM-MC-$M$-DCSK system can significantly increase the transmission rate and improve energy efficiency, and thus has been viewed as a promising physical-layer solution for e-health IoT applications.


To exploit the benefits of both SWIPT and DCSK, an MISO-configured non-coherent short-reference DCSK (SR-DCSK) SWIPT communication system, i.e., SR-DCSK MISO-SWIPT system, has been introduced by using the TS receiver \cite{25}. Although this system can achieve higher data rate than the conventional DCSK system, a part of time slot is employed to send additional chaotic signal for energy harvesting.
To avoid using additional time slot, a CI-DCSK SWPIT system has been developed in \cite{26}. In such a system, the transmitted signals are split into two parts by utilizing TS mode, i.e., the front part of the transmitted signals is utilized in energy harvesting and the later one is adopted in information decoding. However, this system not only changes the receiver structure but also may result in strict time synchronization.
To ensure low-complexity, high-data-rate and  low-power advantages, in this paper we propose a new CIM-MC-$M$-DCSK MISO-SWIPT communication system.\footnote{Because the principles and configures of the CI-DCSK SWIPT and SR-DCSK MISO-SWIPT systems are different from that of the proposed CIM-MC-$M$-DCSK MISO-SWIPT system, it is rather difficult to compare their bit error rates (BERs) with the proposed system. Hence, in this paper, the CIM-MC-$M$-DCSK system is used as the benchmark to validate the superiority of the proposed system.}
The main contributions of this paper are summarized as follows:
\begin{enumerate}
\item
A new MISO-configured SWIPT scheme for the CIM-MC-$M$-DCSK system, i.e., CIM-MC-$M$-DCSK MISO-SWIPT system, is proposed by using orthogonal characteristic of chaotic signals with different initial values, where a multi-antenna is configured at the transmitter to improve energy efficiency while PS receiver is used at the receiver to ensure easy implementation.\footnote{Energy harvesting is significantly affected by the distance between the transmitter and receiver because of the path loss. Hence, multiple antennas are adopted to improve the energy-harvesting efficiency and to achieve high energy efficiency \cite{47}-\cite{52}. However, this paper focuses on the low-complexity and low-power IoT applications. For this reason, we consider the system model with a user equipment (UE) severed by a base station (BS). This system model can be easily extended into multi-user cases due to the broadcast property of wireless signals. In addition,
this system can be easily extended into MIMO scenarios by using equal-gain combining method at the UE with multiple antennas, which does not require channel estimation.} The proposed system is particularly suitable for low-power and low-cost e-health IoT applications, where numerous UEs with the limited battery capacity are served by one base station.

\item
The BER expression of the proposed CIM-MC-$M$-DCSK MISO-SWIPT system is analyzed and derived over multipath Rayleigh fading channels. In particular, an approximation scheme is proposed to derive a closed-form BER expression of $M$-DCSK modulation by using linear processing method, which is also applicable to the other non-coherent chaotic modulations with differential detection, e.g., DCSK \cite{28} and MC-DCSK \cite{30}.
Moreover, simulation results are carried out to validate the theoretical analysis and show the impact of the system parameters on the performance of the proposed system.
Furthermore, the BER results between the proposed system and CIM-MC-$M$-DCSK system are carefully compared to demonstrate the superiority of the proposed system.


\item
Spectral efficiency and energy efficiency of the proposed CIM-MC-$M$-DCSK MISO-SWIPT system are analyzed. It is demonstrated that the proposed system benefits from higher spectral efficiency than the CI-DCSK SWPIT and SR-DCSK MISO-SWIPT systems while it has the same spectral efficiency as that of the CIM-MC-$M$-DCSK system. Moreover, the proposed system can provide higher energy efficiency than that of the CI-DCSK SWPIT, SR-DCSK MISO-SWIPT and CIM-MC-$M$-DCSK systems.




\end{enumerate}

The remainder of this paper is organized as follows. Section II gives the proposed system and signal models. Section III analyzes the performance of the proposed system. Section IV presents various numerical results and discussions. Section V concludes the paper.

\begin{figure}[t] 
\center
\includegraphics[width=3.0in,height=1.2in]{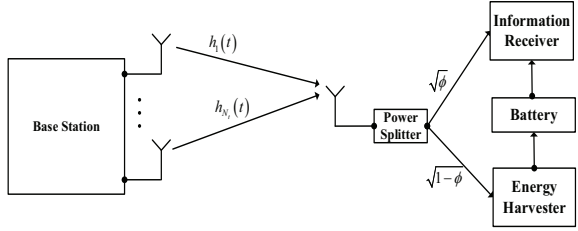}
\caption{An MISO-SWIPT system model.}
\label{fig:Fig.1}
\end{figure}

\begin{figure*}[t]
\center
\subfigure[Transmitter of the $n_t$-th antenna]{ \label{fig:subfig:2a}
\includegraphics[width=6in,height=2.5in]{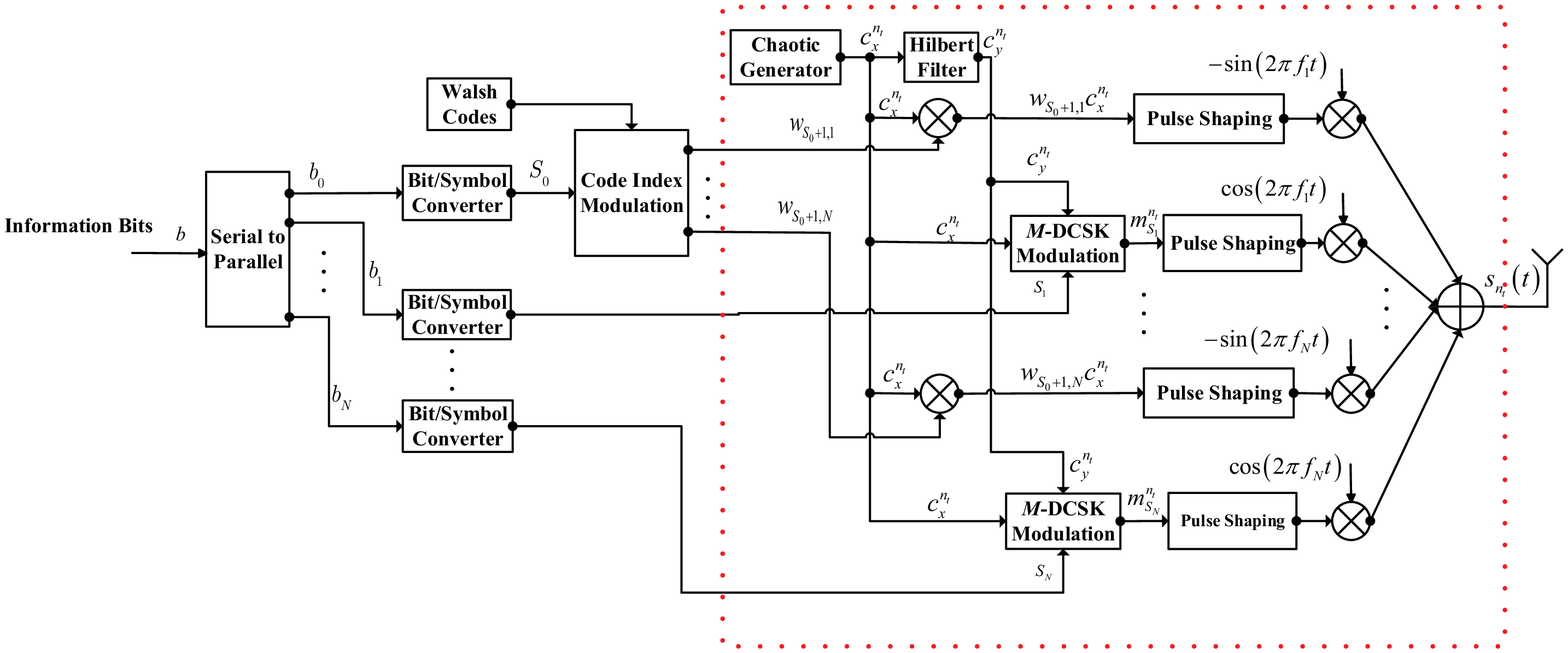}}
\subfigure[Receiver]{ \label{fig:subfig:2b}
\includegraphics[width=6in,height=2.5in]{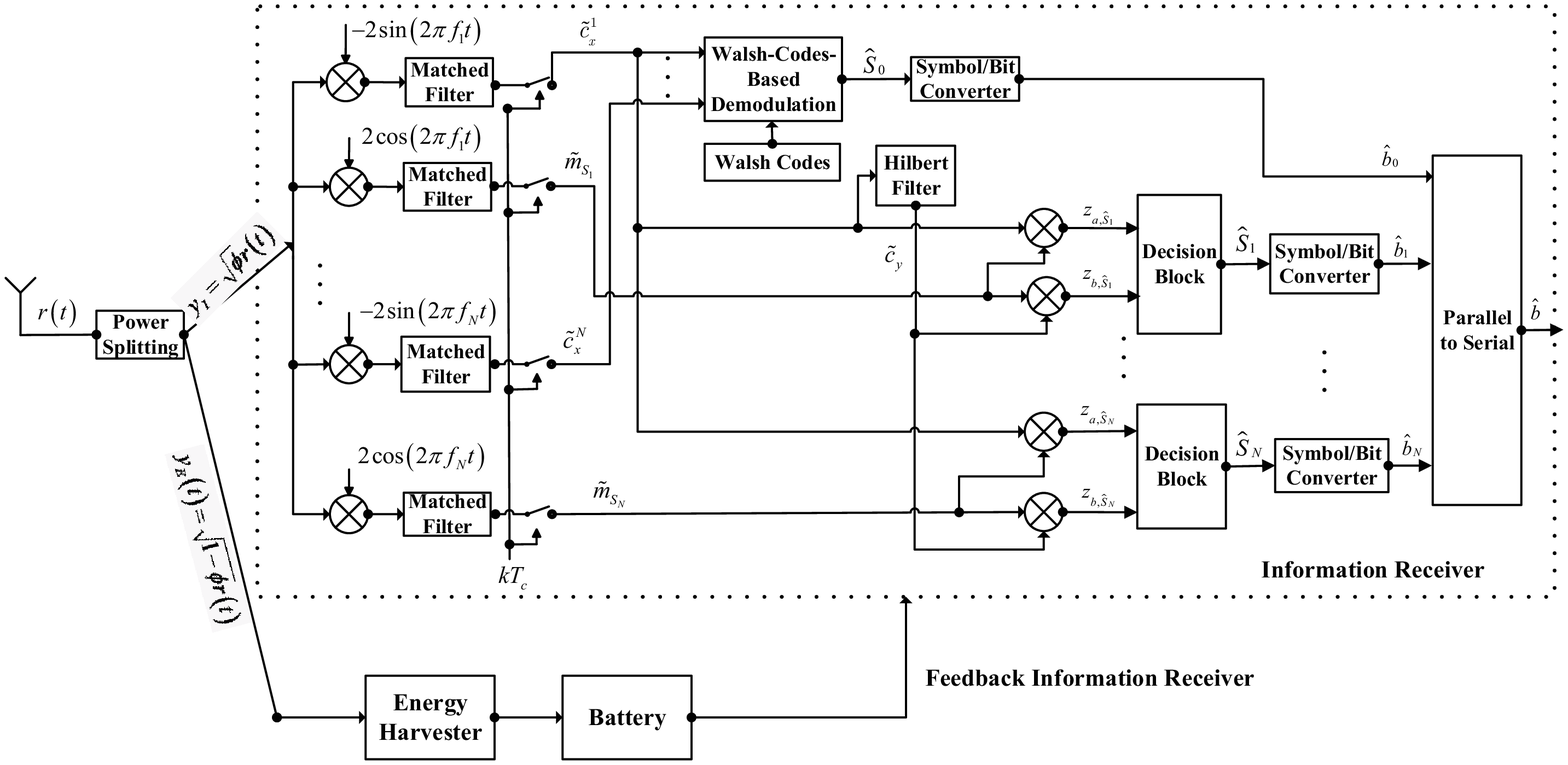}}
\caption{The architecture of a CIM-MC-$M$-DCSK MISO-SWIPT system.}
\label{fig:Fig.2}
\end{figure*}

\section{System and Signal Models}\label{sect:SM}

An MISO-SWIPT system model is considered, which consists of a BS with $N_t$ transmitted antennas
and a UE with a single antenna, as shown in Fig.~\ref{fig:Fig.1},
where the CIM-MC-$M$-DCSK modulation and demodulation shown in Fig.~\ref{fig:Fig.2} are employed at the BS and UE, respectively.
In the considered system, the BS transmits the data information and energy to the UE and after harvesting the energy the UE uses the harvesting energy to decode the information from BS.

\subsection{CIM-MC-DCSK MISO-SWIPT System }\label{sect:PS}

The detail implementation of the CIM-MC-$M$-DCSK MISO-SWIPT system is depicted in Fig.~\ref{fig:Fig.2}, where PS mode is adopted.
At the transmitter of the CIM-MC-$M$-DCSK MISO-SWIPT system as depicted in Fig.~\ref{fig:Fig.2}(a), a Walsh code and $M$-DCSK modulation are adopted in the reference chaotic signal of all the subcarriers and the data chaotic signal of each subcarrier to transmit the data information, respectively. We define the number of the subcarriers for each antenna as $N$, where $N=2^n$ and $n$ is a positive integer.
The CIM-MC-$M$-DCSK MISO-SWIPT system can transmit $N+1$ parallel information, where each parallel information is defined by $b_i$ ($i=0, 1, \ldots, N$). $b_0$ and $b_i$ can be converted into a symbol $S_0$ ($S_0 \in \{0, 1, \ldots, N-1 \}$) and a symbol $S_i$ ($S_i \in \{0, 1, \ldots, M-1 \}$), respectively. Hence, in the designed CIM-MC-$M$-DCSK MISO-SWIPT system, it transmits the same information data for each antenna.

\begin{figure}[t] 
\center
\includegraphics[width=3.5in,height=2.6in]{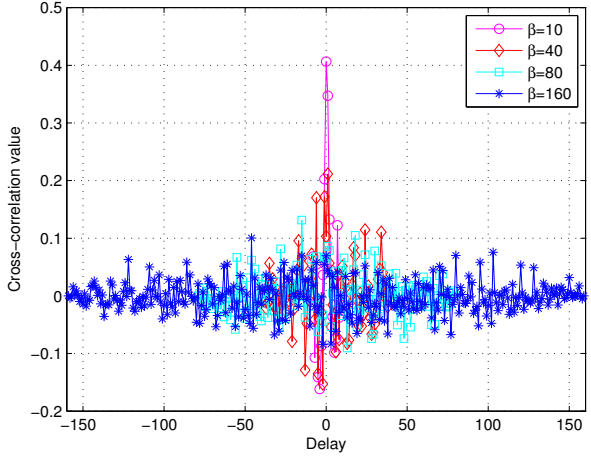}
\caption{Cross-correlation values of two chaotic signals with different initial condition values.}
\label{fig:Fig.3-1}
\end{figure}

\begin{figure}[t] 
\center
\includegraphics[width=3.2in,height=3.2in]{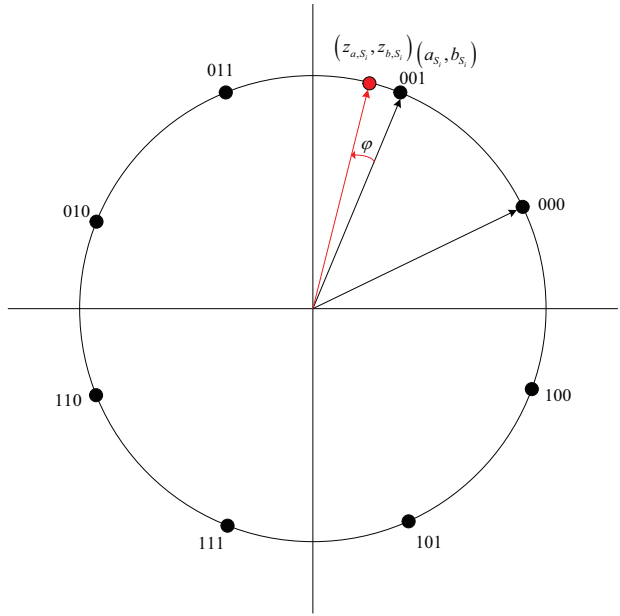}
\vspace{-0.1 cm}
\caption{An $8$-DCSK constellation.}
\label{fig:Fig.3}
\end{figure}

To mitigate the interference between the antennas, different chaotic signals are used at each antenna.
The chaotic signals generated with different initial condition values have good cross correlation \cite{52-1,52-2}, which tends to zero for a large spreading factor.
As a simple example, Fig.~\ref{fig:Fig.3-1} presents the cross-correlation values of two chaotic signals with different initial condition values, where a logistic map, i.e., $x_{k+1} = 1-2x_{k}^2$, is adopted. Hence, one can easily obtain infinite number of the orthogonal chaotic signals.
The chaotic signal used in the $n_t$-th antenna is defined as $c_{x}^{n_t}=(c_{x,1}^{n_t},\ldots,c_{x,\beta}^{n_t})$, where $\beta$ presents the spreading factor and $n_t = 1, \ldots, N_t$.
For a CIM, the in-phase baseband signal can be given by $w_{S_0+1}^{T}c_{x}^{n_t}$, where $w_{S_0+1}$ denotes the $(S_0+1)$-th row of the Walsh-code matrix and $(\cdot)^T$ presents the transpose operation.
For an $M$-DCSK modulation, the quadrature baseband signal can be created by a linear combination of
chaotic signals, i.e., $m_{S_{i}}^{n_t}=a_{S_i}c_{x}^{n_t} +b_{S_i}c_{y}^{n_t}$, where $c_{y}^{n_t}$ is the Hilbert transform form of $c_{x}^{n_t}$, $a_s$ and $b_s$ present the $x$-axis and $y$-axis coordinate
values corresponding to a constellation point, e.g., an $8$-DCSK constellation shown in Fig.~\ref{fig:Fig.3}.
The in-phase signal and quadrature signal are written as $x_{I,i}^{n_t}(t)=\sum_{k=1}^{\beta}m_{S_i,k}^{n_t}q(t-kT_c)$ and $x_{Q,i}^{n_t}(t)=\sum_{k=1}^{\beta}w_{S_0+1,i}c_{x,k}^{n_t}q(t-kT_c)$, respectively, where $T_c$ denotes chip time and $q(t)$ is the pulse shaping with a square-root-raised-cosine filter.
Therefore, the transmitted signal of the $n_t$-th antenna for the CIM-MC-$M$-DCSK MISO-SWIPT system can be given by
\begin{align}
\hspace{-1.5mm} s_{n_t}(t)= \frac{1}{\sqrt{N_t}} \sum_{i=1}^{N} \Big{(} x_{I,i}^{n_t}(t)\cos \left( 2\pi f_it \right) \hspace{-1mm} - \hspace{-1mm} x_{Q,i}^{n_t}(t)\sin \left( 2\pi f_it \right) \Big{)},
\label{eq:Eq_1}
\end{align}
where $f_i$ denotes the frequency of the sinusoidal or cosine carrier and it satisfies $f_i \gg 1/T_c$, and  $f_i$ and $f_j$ ($i\ne j$) have mutual orthogonality.

After the multipath fading channels $h_{n_t}(t)$ $(n_t=1,...,N)$, the received signal of the user is given by
\begin{align}
r(t) = \sum_{n_t=1}^{N_t}\sum_{l=1}^{L_{n_t}}\alpha_{l,n_t} s_{n_t}(t-\tau_{l,n_t})+n(t),
\label{eq:Eq_2}
\end{align}
where $n(t)$ is the additive white Gaussian noise (AWGN) with zero mean and variance of $\frac{N_0}{2}$, $L_{n_t}$ is the number of paths, $\alpha_{l,n_t}$ and  $\tau_{l,n_t}$ present the channel coefficient and the path delay of the $l$-th path, respectively.

As shown in Fig.~\ref{fig:Fig.2}(b), $r(t)$ is divided into two parts:
$y_{E}(t)=\sqrt{1-\phi}r(t)$ is utilized for harvesting energy, and $y_{I}(t)=\sqrt{\phi}r(t)$ is adopted to demodulate information, where the parameter $\phi$ denotes the power-splitting ratio and $0 \le \phi \le 1$.
Then, $y_{I}(t)$ is handled by the cosine and sinusoidal carriers as well as the matched filters. Hence, one gets the in-phase and quadrature baseband signals of the $i$-th subcarrier, which are respectively given by
\begin{align}
\hspace{-1.5mm} \tilde{c}_{x}^{i} =  \sum_{k=1}^{\beta} \Bigg{(} \sqrt{\phi} \sum_{n_t=1}^{N_t} \sum_{l=1}^{L_{n_t}} \frac{\alpha_{l,n_t} }{\sqrt{N_t}}  w_{S_0+1,i}c_{x,k-\tau_{l,n_t}}^{n_t} + n_{x,k}^{i} \Bigg{)},
\label{eq:Eq_3}
\end{align}
\vspace{-4.5mm}
\begin{align}
\hspace{-1.5mm}\tilde{m}_{S_i} =&  \sum_{k=1}^{\beta} \Bigg{(} \sqrt{\phi}  \sum_{n_t=1}^{N_t} \sum_{l=1}^{L_{n_t}}  \frac{\alpha_{l,n_t} }{\sqrt{N_t}} \left( a_{S_i}c_{x,k-\tau_{l,n_t}}^{n_t} \hspace{-1mm} + \hspace{-1mm} b_{S_i}c_{y,k-\tau_{l,n_t}}^{n_t} \right) \nonumber \\
 & + n_{x,k+\beta}^{i}  \Bigg{)},
\label{eq:Eq_4}
\end{align}
where $n_{x,k}^{i}$ denotes the baseband noise of the $i$-th subcarrier.\footnote{Generally, the antenna noise can be ignored \cite{5,9}, thus in this paper we do not consider this part of the noise.}
Finally, according to the demodulation principle in \cite{46}, the in-phase and quadrature baseband signals are deployed to decode the information from BS.

\subsection{Energy Harvester}
The receiver harvests the energy from the signal $y_{E}(t)=\sqrt{1-\phi}r(t)$ by performing a RF-to-DC conversion. The linear energy harvesting model is considered.
Through some simple mathematical calculations for $y_{E}(t)$, the harvested power at the UE can be obtained as
\begin{align}
P_{h} = \frac{2NE_1\lambda(1-\phi)}{N_t} \sum_{n_t=1}^{N_t} \sum_{l=1}^{L_{n_t}}\alpha_{l,n_t}^2,
\label{eq:Eq_5}
\end{align}
where $E_1=\sum_{k=1}^{\beta}c_{x,k}^2$ is the energy of a chaotic signal $c_{x}$ and $\lambda$ denotes the energy conversion efficiency factor satisfying $0 \le \lambda \le 1$.



\section{Performance Analysis}\label{sect:PAMR}
In this section, BER expressions for the proposed system are first derived over multipath fading channels. Then, energy efficiency and spectral efficiency for the proposed system are analyzed.



\subsection{BER Analysis}

In this paper, it is assumed that the largest multipath delay $\tau_{L_{n_t}}$ of the $n_t$-th antenna is much shorter than the symbol duration, i.e., $0<\tau_{L_{n_t}}\ll \beta$, thus the inter-symbol interference (ISI) can be negligible \cite{8,10,19,13}. Moreover, it is also assumed that the channel from each antenna to the UE is independent block fading and quasi-static during a symbol duration. Without loss of generality, we also assume that the gains of the channels from the BS to the UE are different, i.e., $\alpha_{1,1} \ne \ldots \ne \alpha_{1,L_{n_t}} \ne \ldots \ne \alpha_{N_t,L_{N_t}}$, which are independent Rayleigh distribution random variable.
Moreover, we assume that at the information receiver for the CIM-MC-$M$-DCSK MISO-SWIPT system  any external power supply is not used and it only needs a power $P_R$ to recover the transmitted information data.\footnote{In this paper, we only consider a SWIPT communication system, where an energy-harvesting UE harvests energy and decodes information from a BS in the downlink. However, in practical IoT applications, the UE can use the remaining harvested energy to transmit data to the BS in the uplink.}
According to the definition on the energy shortage probability in \cite{25}, when the harvested power is less than the power $P_R$, the transmitted information data can not be recovered, thus the energy shortage occurs. Hence, the system BER $P_{sys}$ of the proposed CIM-MC-$M$-DCSK MISO-SWIPT system is a function of the energy shortage probability $P_{Shr}$ and the BER $P_{b}$ of the information receiver.
Furthermore, the BER $P_{b}$ of the information receiver includes two parts: the BER $P_{b,CIM}$ of the CIM and the BER $P_{b,MDCSK}$ of the $M$-DCSK modulation. In addition, the number of the transmitted bits for the proposed system is $\log_2N+N\log_2M$. The average energy per bit, i.e., $E_b$, is denoted by $E_b = \frac{2NE_1}{\log_2N+N\log_2M}$.
Therefore, the system BER of the CIM-MC-$M$-DCSK MISO-SWIPT system can be expressed by
\begin{align}
P_{sys} = &(1-P_{Shr})P_b + P_{Shr} \nonumber \\
=&\left(1-P_{Shr}\right)\times \left(\frac{\log_2N}{\log_2N+N\log_2M}P_{b,CIM}+ \right. \nonumber \\
&\left. \frac{N\log_2M}{\log_2N+N\log_2M}P_{b,MDCSK}\right) + P_{Shr}.
\label{eq:Eq_7}
\end{align}
In the following, the energy shortage probability $P_{Shr}$, BER of the CIM $P_{b,CIM}$ and BER of the $M$-DCSK $P_{b,MDCSK}$ are derived.

\subsubsection{Energy Shortage Probability $P_{Shr}$}

The expression of the energy shortage probability is given by
\begin{align}
P_{Shr} &= \Pr\{P_h<P_R\} \nonumber \\
&=\Pr \Big{\{} \frac{2NE_1\lambda(1-\phi)}{N_t} \sum_{n_t=1}^{N_t} \sum_{l=1}^{L_{n_t}}\alpha_{l,n_t}^2 < P_R \Big{\}} \nonumber \\
&=\Pr \Big{\{}  \frac{1}{N_t} \sum_{n_t=1}^{N_t} \sum_{l=1}^{L_{n_t}}\alpha_{l,n_t}^2 < \frac{ P_R}{2NE_1\lambda(1-\phi)} \Big{\}}.
\label{eq:Eq_8}
\end{align}

Since the variable $\alpha_{l,n_t}$ follows the Rayleigh distribution, $\alpha_{l,n_t}^2$ is a chi-square distribution with $2$ degrees of freedom (i.e., $\alpha_{l,n_t}^2 \sim \chi_{2}^2$).
The variable $\frac{1}{N_t} \sum_{n_t=1}^{N_t} \sum_{l=1}^{L_{n_t}}\alpha_{l,n_t}^2$ should be decomposed into the sum of $\sum_{n_t=1}^{N_t} L_{n_t}$ variables, where each variable follows the chi-square distribution of $\chi_{2}^2$. The characteristics function in \cite{53} can be used to calculate the probability density function (PDF) of the variable $ \frac{1}{N_t} \sum_{n_t=1}^{N_t} \sum_{l=1}^{L_{n_t}}\alpha_{l,n_t}^2$, given by
\begin{flalign}
f(x)= \sum_{n_t=1}^{N_t} \sum_{l=1}^{L_{n_t}} \frac{\nu_{l,n_t}}{ \bar{x}_{l,n_t}} \exp \left( -\frac{x}{\bar{x}_{l,n_t}} \right),
\label{eq:Eq_8-1}
\end{flalign}
where $\bar{x}_{l,n_t}$ is defined as
\begin{flalign}
\bar{x}_{l,n_t} = \frac{1}{N_t} \E \left( \alpha_{l,n_t}^2 \right),
\label{eq:Eq_8-2}
\end{flalign}
and $\nu_{l,n_t}$ is defined as
\begin{flalign}
\nu_{l,n_t} = \prod_{i=1,i \ne l}^{L_{n_t}} \frac{\bar{x}_{l,n_t}}{ \bar{x}_{l,n_t} - \bar{x}_{i,n_t}} \prod_{n=1, n \ne n_t}^{N_t} \prod_{j=1}^{L_{n}} \frac{\bar{x}_{l,n_t}}{ \bar{x}_{l,n_t} - \bar{x}_{j,n}}.
\label{eq:Eq_8-3}
\end{flalign}

Hence, Eq.~\eqref{eq:Eq_8} can be computed as
\begin{align}
P_{Shr} &= \sum_{n_t=1}^{N_t} \sum_{l=1}^{L_{n_t}} \frac{\nu_{l,n_t}}{ \bar{x}_{l,n_t}} \int_{0}^{\frac{ P_R}{2NE_1\lambda(1-\phi)}} \exp \left( -\frac{x}{\bar{x}_{l,n_t}} \right) dx \nonumber \\
&=  \sum_{n_t=1}^{N_t} \sum_{l=1}^{L_{n_t}} \nu_{l,n_t} \left( 1- \exp \left( -\frac{ P_R}{2NE_1\bar{x}_{l,n_t}\lambda(1-\phi)} \right) \right).
\label{eq:Eq_8-4}
\end{align}

\subsubsection{BER of the CIM $P_{b,CIM}$}

The $j$-th~$(j=1,2,\ldots,N)$ energy metric of the demodulation for the CIM can be written as
\begin{align}
z_{e,j} =  \sum_{k=1}^{\beta} \Bigg{[}\sum_{i=1}^{N} \Bigg{(} \sqrt{\phi} \sum_{n_t=1}^{N_t} \sum_{l=1}^{L_{n_t}} \frac{\alpha_{l,n_t} }{\sqrt{N_t}} & w_{j,i} w_{S_0+1,i}c_{x,k-\tau_{l,n_t}}^{n_t} \nonumber \\
& + w_{j,i}n_{x,k}^{i} \Bigg{)} \Bigg{]}^2,
\label{eq:Eq_12}
\end{align}

When $j = S_0 +1$ and considering the orthogonality of chaotic signal, the energy metric is calculated as
\begin{align}
z_{e,j} =  \sum_{k=1}^{\beta} \Bigg{(} \sqrt{\phi} N \sum_{n_t=1}^{N_t} \sum_{l=1}^{L_{n_t}} \frac{\alpha_{l,n_t} }{\sqrt{N_t}}  c_{x,k-\tau_{l,n_t}}^{n_t} \hspace{-1mm} + \hspace{-1mm} \sum_{i=1}^{N}w_{j,i}n_{x,k}^{i} \Bigg{)}^2.
\label{eq:Eq_13}
\end{align}

The mean and variance of Eq.~\eqref{eq:Eq_13} are computed as, respectively,
\begin{align}
\E[z_{e,j}] = \mu_1 = \frac{\phi N^2 E_1}{N_t}  \sum_{n_t=1}^{N_t} \sum_{l=1}^{L_{n_t}} \alpha_{l,n_t}^2 + \frac{\beta NN_0}{2},
\label{eq:Eq_14}
\end{align}
\vspace{-3.5mm}
\begin{align}
Var[z_{e,j}] = \sigma_{1}^{2} = \frac{2\phi N^3 E_1 N_0}{N_t}  \sum_{n_t=1}^{N_t} \sum_{l=1}^{L_{n_t}} \alpha_{l,n_t}^2 + \frac{\beta N^2 N_{0}^{2}}{2},
\label{eq:Eq_15}
\end{align}
where $\E[\bullet]$ denotes the expectation operator and $Var[\bullet]$ is the variance operator.
The detailed calculations of Eqs.~\eqref{eq:Eq_14} and \eqref{eq:Eq_15} can be found in Appendix A.

When $j\neq S_0 +1$, the energy metric is calculated as
\begin{align}
z_{e,j} = \sum_{k=1}^{\beta}  \left( \sum_{i=1}^{N}w_{j,i}n_{x,k}^{i} \right)^2.
\label{eq:Eq_16}
\end{align}
The mean and variance of Eq.~\eqref{eq:Eq_16} can be computed as, respectively,
\begin{align}
\E[z_{e,j}] = \mu_2 = \frac{\beta NN_0}{2}, Var[z_{e,j}] = \sigma_{2}^2= \frac{\beta N^2 N_{0}^{2}}{2}.
\label{eq:Eq_17}
\end{align}
The detailed calculations of Eq.~\eqref{eq:Eq_17} can be found in Appendix A.

Hence, all the elements of the energy metric vector $z_e = [z_{e,1},\ldots,z_{e,j},\dots,z_{e,N}]$ are mutually independent, which follow the Gaussian distribution.
The symbol error rate (SER) expression of the CIM can be derived as
\begin{flalign}
&P_{s,CIM}(\gamma_b)=\nonumber\\
&  1-\int_{-\infty}^{\infty}
\left( \int_{-\infty}^{z_{e,j}}
\frac{1}{\sqrt{2 \pi \sigma_{2}^2}} \exp\left( - \frac{\left(z_{e,i}-\mu_2\right)^2}{2\sigma_{2}^2} \right)
dz_{e,i}\right)^{N-1}  \nonumber\\
& ~\times \frac{1}{\sqrt{2 \pi \sigma_{1}^2}} \exp\left( - \frac{\left(z_{e,j}-\mu_1\right)^2}{2\sigma_{1}^2} \right) dz_{e,j},
\label{eq:Eq_18}
\end{flalign}
where $\gamma_b=\sum_{n_t=1}^{N_t} \sum_{l=1}^{L_{n_t}} \frac{\alpha_{l,n_t}^2 }{N_t} (E_b/N_0)$.

Let $t=z_{e,i}/\sigma_2$ for $i \neq j$ and $u=z_{e,j}/\sigma_2$,  Eq.~\eqref{eq:Eq_8} can be further simplified to
\begin{flalign}
P_{s,CIM}(\gamma_b) = &  \frac{\eta_1}{\sqrt{2 \pi}} \int_{-\infty}^{\infty}
\left( 1-\left( 1-Q(u)\right)^{N-1} \right) \nonumber\\
  & \times \exp \left( - \frac{\left(\eta_1 \mu-\gamma_1 \right)^2}{2} \right)  du,
\label{eq:Eq_19}
\end{flalign}
where $Q(u)=\frac{1}{\sqrt{2 \pi}}\int_{x}^{\infty} \exp (-\frac{t^2}{2}) dt$.
Using Eqs.~\eqref{eq:Eq_14} and \eqref{eq:Eq_15}, one can further obtain
\begin{flalign}
\gamma_1 = &\frac{\mu_1-\mu_2}{\sigma_1} = \frac{\phi (\log_2N+N\log_2M)\gamma_b}{\sqrt{4\phi (\log_2N+N\log_2M)\gamma_b + 2\beta } }, \nonumber \\
\eta_1 = &\frac{\sigma_2}{\sigma_1} = \frac{1}{ \sqrt{\frac{2\phi (\log_2N+N\log_2M)}{\beta}\gamma_b + 1 } }.
\label{eq:Eq_20-1}
\end{flalign}


Using $Q(x) \le \frac{1}{2}\exp(-\frac{x^2}{2})$ and the method of the Taylor series expansion, a closed-form expression of Eq.~\eqref{eq:Eq_19} can be obtained, i.e., Eq.~\eqref{eq:Eq_21}, where $\dbinom{n}{m}$ is the combinational number.
\begin{figure*}[t]
\hrulefill
\begin{flalign}
P_{s,CIM}(\gamma_b) \approx &     \frac{\eta_1}{\sqrt{2 \pi}} \int_{-\infty}^{\infty}
\left( 1-\left( 1-\frac{1}{2}\exp\left(-\frac{x^2}{2}\right)\right)^{N-1} \right) \times \exp \left( - \frac{\left(\eta_1\mu-\gamma_1 \right)^2}{2} \right)  du \nonumber \\
= & \frac{\eta_1}{\sqrt{2 \pi}} \int_{-\infty}^{\infty} \left( 1-\sum_{n=0}^{N-1}\left(\frac{-1}{2}\right)^n \dbinom{N-1}{n} \exp\left( -\frac{n\mu^2}{2}\right) \right) \times \exp \left( - \frac{\left(\eta_1\mu-\gamma_1 \right)^2}{2} \right)  du \nonumber \\
= & \sum_{n=1}^{N-1}\left(\frac{-1}{2}\right)^{n+1}  \frac{ \dbinom{N-1}{n} \eta_1 }{n+\eta_1}\exp \left( -\frac{(n+\eta_1-\eta_{1}^{2})\gamma_{1}^{2}}{2(n+\eta_1)} \right).
\label{eq:Eq_21}
\end{flalign}
\end{figure*}

Hence, the BER expression of the CIM can be obtained as
\begin{flalign}
P_{b,CIM}(\gamma_b) = \frac{N}{2(N-1)}P_{s,CIM}(\gamma_b).
\label{eq:Eq_22}
\end{flalign}

Using the characteristics function,  $\gamma_b$ can be calculated as
\begin{flalign}
f(\gamma_b)= \sum_{n_t=1}^{N_t} \sum_{l=1}^{L_{n_t}} \frac{\pi_{l,n_t}}{ \bar{\gamma}_{l,n_t}} \exp \left( -\frac{\gamma_b}{\bar{\gamma}_{l,n_t}} \right),
\label{eq:Eq_23}
\end{flalign}
where $\bar{\gamma}_{l,n_t}$ is the average bit-SNR for the $l$-th path of the $n_t$-th antenna, defined as
\begin{flalign}
\bar{\gamma}_{l,n_t} = \frac{1}{N_t} \frac{E_b}{N_0}\E \left(\alpha_{l,n_t}^2 \right),
\label{eq:Eq_24}
\end{flalign}
and $\pi_{l,n_t}$ is defined as
\begin{flalign}
\pi_{l,n_t} = \prod_{i=1,i \ne l}^{L_{n_t}} \frac{\bar{\gamma}_{l,n_t}}{ \bar{\gamma}_{l,n_t} - \bar{\gamma}_{i,n_t}} \prod_{n=1,n \ne n_t}^{N_t} \prod_{j=1}^{L_n} \frac{\bar{\gamma}_{l,n_t}}{ \bar{\gamma}_{l,n_t} - \bar{\gamma}_{j,n}}.
\label{eq:Eq_25}
\end{flalign}
Therefore, combining Eq.~\eqref{eq:Eq_23} with Eq.~\eqref{eq:Eq_22}, one can obtain $P_{b,CIM}$, i.e., Eq.~(\ref{eq:Eq_27}). Although a closed-form expression cannot be obtained, Eq.~(\ref{eq:Eq_27}) can be easily computed numerically.
\begin{figure*}[t]
\hrulefill
\begin{flalign}
P_{b,CIM} = \frac{N}{2(N-1)} \sum_{n_t=1}^{N_t} \sum_{l=1}^{L_{n_t}} \frac{\pi_{l,n_t}}{ \bar{\gamma}_{l,n_t}} \sum_{n=1}^{N-1}\left(\frac{-1}{2}\right)^{n+1} \dbinom{N-1}{n}  \int_{0}^{\infty} \frac{\eta_1}{n+\eta_1} \exp \left( -\frac{(n+\eta_1-\eta_{1}^{2})\gamma_{1}^{2}}{2(n+\eta_1)} -\frac{\gamma_b}{\bar{\gamma}_{l,n_t}} \right) d \gamma_b.
\label{eq:Eq_27}
\end{flalign}
\end{figure*}





\subsubsection{BER of the $M$-DCSK $P_{b,MDCSK}$}
Similar to the demodulation principle of $M$-DCSK part in \cite{46}, the decision variables at the output of the correlators of the $i$-th subcarrier can be approximated as, respectively,
\begin{align}
&z_{a,\hat{S}_i} \approx \sum_{k=1}^{\beta} \left( \sqrt{\phi} \sum_{n_t=1}^{N_t} \sum_{l=1}^{L_{n_t}} \frac{\alpha_{l,n_t} }{\sqrt{N_t}} c_{x,k-\tau_{l,n_t}}^{n_t} + n_{x,k}^{1} \right) \times \nonumber \\
\hspace{-1mm}& \hspace{-1mm}\left( \sqrt{\phi}  \sum_{n_t=1}^{N_t} \sum_{l=1}^{L_{n_t}}  \frac{\alpha_{l,n_t} }{\sqrt{N_t}} \hspace{-1mm} \left(a_{S_i}c_{x,k-\tau_{l,n_t}}^{n_t} \hspace{-1mm} + \hspace{-1mm}  b_{S_i}c_{y,k-\tau_{l,n_t}}^{n_t} \hspace{-1mm} \right) \hspace{-1mm} + \hspace{-1mm} n_{x,k+\beta}^{i}  \hspace{-1mm} \right),
\label{eq:Eq_28}
\end{align}
\vspace{-3.5mm}
\begin{align}
&z_{b,\hat{S}_i} \approx \sum_{k=1}^{\beta} \left( \sqrt{\phi} \sum_{n_t=1}^{N_t} \sum_{l=1}^{L_{n_t}} \frac{\alpha_{l,n_t} }{\sqrt{N_t}} c_{y,k-\tau_{l,n_t}}^{n_t} + \tilde{n}_{x,k}^{1} \right) \times \nonumber \\
& \hspace{-1mm} \left( \sqrt{\phi}  \sum_{n_t=1}^{N_t} \sum_{l=1}^{L_{n_t}}  \frac{\alpha_{l,n_t} }{\sqrt{N_t}} \hspace{-1mm} \left(a_{S_i}c_{x,k-\tau_{l,n_t}}^{n_t} \hspace{-1mm} + \hspace{-1mm} b_{S_i}c_{y,k-\tau_{l,n_t}}^{n_t} \hspace{-1mm} \right) \hspace{-1mm} + \hspace{-1mm} n_{x,k+\beta}^{i}  \hspace{-1mm} \right),
\label{eq:Eq_29}
\end{align}
where $ \tilde{n}_{x,k}^{1} $ denotes the Hilbert transform of $n_{x,k}^{1}$ and presents an AWGN with
zero mean and variance $\frac{N_0}{2}$.

The means and the variances of Eqs.~\eqref{eq:Eq_28} and \eqref{eq:Eq_29} can be given by
\begin{align}
\E[z_{a,\hat{S}_i}] = \frac{\phi a_{S_i} E_1}{N_t}  \sum_{n_t=1}^{N_t} \sum_{l=1}^{L_{n_t}} \alpha_{l,n_t}^2,
\label{eq:Eq_30}
\end{align}
\begin{align}
\E[z_{b,\hat{S}_i}] = \frac{\phi b_{S_i} E_1}{N_t}  \sum_{n_t=1}^{N_t} \sum_{l=1}^{L_{n_t}} \alpha_{l,n_t}^2,
\label{eq:Eq_31}
\end{align}
\begin{align}
Var[z_{a,\hat{S}_i}] = Var[z_{b,\hat{S}_i}] = \frac{\phi E_1 N_0}{N_t}  \sum_{n_t=1}^{N_t} \sum_{l=1}^{L_{n_t}} \alpha_{l,n_t}^2 + \frac{\beta N_{0}^{2}}{4}.
\label{eq:Eq_32}
\end{align}
The detailed calculations of Eqs.~\eqref{eq:Eq_30}, \eqref{eq:Eq_31} and \eqref{eq:Eq_32}  can be found in Appendix B.

Through the polar coordinates transformations of $z_{a,\hat{S}_i}$ and $z_{b,\hat{S}_i}$, the instantaneous PDF can be computed as
\begin{flalign}
p\left(\varphi|\gamma_b\right)=&\frac{1}{2\pi}\exp\left(-\frac{\gamma^2}{8}\right)+\frac{\gamma \cos \varphi}{2 \sqrt{2\pi}} \times \exp\left(-\frac{\gamma^2 \sin^2\varphi}{8}\right) \nonumber\\
&\times Q\left(-\frac{\gamma \cos \varphi}{2}\right),
\label{eq:Eq_33}
\end{flalign}
where $\varphi$, as shown in Fig.~\ref{fig:Fig.3}, presents the phase error between the transmitted constellation and the received one, and the parameter $\gamma$ is calculated as
\begin{flalign}
\gamma=\frac{2 \phi (\log_2N+N\log_2M) \gamma_b}{ \sqrt{2 \phi N (\log_2N+N\log_2M) \gamma_b+ N^2 \beta} }.
\label{eq:Eq_34}
\end{flalign}

For $M$-DCSK modulation, by using the PDF in Eq.~\eqref{eq:Eq_33}, the BER expression can be obtained as
\begin{flalign}
P_{b,MDCSK}(\gamma_b)=  \frac{1}{\log_2M} \left( 1- \int_{-\frac{\pi}{M}}^{\frac{\pi}{M}} p\left(\varphi|\gamma_b\right)d \varphi \right).
\label{eq:Eq_35}
\end{flalign}
According to \cite{46}, at high SNR regime, the approximated expression of Eq.~\eqref{eq:Eq_35} is given by
\begin{flalign}
P_{b,MDCSK}(\gamma_b)  \approx \frac{2}{\log_2M} Q\left(\frac{\gamma \sin (\frac{\pi}{M})}{2}\right).
\label{eq:Eq_36}
\end{flalign}
Hence, combining Eq.~\eqref{eq:Eq_23} with Eq.~\eqref{eq:Eq_36}, one can get $P_{b,MDCSK}$, i.e., Eq.~(\ref{eq:Eq_38}).
\begin{figure*}[t]
\hrulefill
\begin{flalign}
P_{b,MDCSK}  \approx  \frac{2}{\log_2M} \sum_{n_t=1}^{N_t} \sum_{l=1}^{L_{n_t}} \frac{\pi_{l,n_t}}{ \bar{\gamma}_{l,n_t}} \int_{0}^{\infty} Q\left(\frac{ \phi (\log_2N+N\log_2M) \sin (\frac{\pi}{M}) \gamma_b}{2\sqrt{ \phi N (\log_2N+N\log_2M) \gamma_b+ N^2 \beta}}\right) \exp \left( -\frac{\gamma_b}{\bar{\gamma}_{l,n_t}} \right) d \gamma_b.
\label{eq:Eq_38}
\end{flalign}
\end{figure*}

To obtain closed-form expression for Eq.~\eqref{eq:Eq_38}, we develop an approximation scheme by using the linear process, given by
\begin{flalign}
Q(\frac{Ax}{\sqrt{Bx+C}}) \simeq \Xi (A,B,C), \nonumber \\
\Xi (A,B,C) = kx+\frac{1}{2}, 0 \leq x \leq -\frac{1}{2k},
\label{eq:Eq_38-1}
\end{flalign}
where $A = \phi (\log_2N+N\log_2M) \sin (\frac{\pi}{M})$, $B=4 \phi N (\log_2N+N\log_2M)$, $C=4N^2 \beta$, $k= \frac{y_0-\frac{1}{2}}{x_0}$, $y_0 = Q(\frac{Ax_0}{\sqrt{Bx_0+C}} )$.
Here, the derivative of $Q(\frac{Ax}{\sqrt{Bx+C}} )$ is computed and $x = 0$ is took into the result of derivation, then one has $-\frac{A}{2N\sqrt{2\pi \beta}}$.
Taking $(0,\frac{1}{2})$ as the tangent point and $-\frac{A}{2N\sqrt{2\pi \beta}}$ as the slope, a linear equation can be obtained as $p = -\frac{A}{2N\sqrt{2\pi \beta}}q + \frac{1}{2}$. Let $p= 0$, then one has $q = \frac{A}{4N\sqrt{2\pi \beta}}$.
Let $x_0 = q$, one can obtain $y_0$, thus $k$ can be computed as
\begin{flalign}
k = \frac{Q\left( \frac{\frac{A^2}{4N\sqrt{2\pi \beta}}}{\sqrt{\frac{AB}{4N\sqrt{2\pi \beta}}+ C} }\right) - \frac{1}{2}}{\frac{A}{4N\sqrt{2\pi \beta}}}.
\label{eq:Eq_38-2}
\end{flalign}
%
Hence, combining Eq.~\eqref{eq:Eq_38-1} with Eq.~\eqref{eq:Eq_38}, one can obtain a closed-form expression, i.e., Eq~\eqref{eq:Eq_38-3}.
\begin{figure*}[t]
\hrulefill
\begin{flalign}
P_{b,MDCSK}  &\approx  \frac{2}{\log_2M} \sum_{n_t=1}^{N_t} \sum_{l=1}^{L_{n_t}} \frac{\pi_{l,n_t}}{ \bar{\gamma}_{l,n_t}} \int_{0}^{\infty} (k \gamma_b +\frac{1}{2}) \exp \left( -\frac{\gamma_b}{\bar{\gamma}_{l,n_t}} \right) d \gamma_b \nonumber \\
&= \frac{2}{\log_2M} \sum_{n_t=1}^{N_t} \sum_{l=1}^{L_{n_t}} \pi_{l,n_t}
\left( k \bar{\gamma}_{l,n_t} \left(1-\exp \left(\frac{1}{2k\bar{\gamma}_{l,n_t}}\right) \right) + \frac{1}{2}\right).
\label{eq:Eq_38-3}
\end{flalign}
\end{figure*}
It should be noted that the proposed approximation scheme can be used for the closed-form BER derivation of the other non-coherent chaotic modulations with differential detection, e.g., DCSK and MC-DCSK modulation.
\\{{\em Remark:} Actually, the closed-form BER expression (i.e., Eq.~\eqref{eq:Eq_38-3}) of the $M$-DCSK system is the lower bound of the integral BER expression (i.e., Eq.~\eqref{eq:Eq_38}), which is used for the diversity analysis.
Through the simulations, it can be found that in the low  SNR regime the results of Eq.~\eqref{eq:Eq_38-3} can well match with the ones of Eq.~\eqref{eq:Eq_38} while in the high SNR regime there is about 2~dB gap. However, the comparison results are not given in this paper due to the space limitation. In the future, some new methods will be explored to derive more accurate closed-form BER of the $M$-DCSK system.}

\subsection{Spectral Efficiency Analysis}
Spectral efficiency (b/s/Hz) is defined as the ratio between the transmitted bits and the number of subcarriers \cite{54}. Hence, the spectral efficiency of the proposed CIM-MC-$M$-DCSK MISO-SWIPT system is given by
\begin{flalign}
SE_{P} =  \frac{\log_2N+N\log_2M}{N}.
\label{eq:SE}
\end{flalign}

The spectral efficiency of the CI-DCSK SWIPT system is $SE_{CI}=\frac{\log_2N+N-1}{N+1}$ b/s/Hz, while the spectral efficiency of the SR-DCSK MISO-SWIPT system is about $1$ b/s/Hz.
It is evident that the CIM-MC-$M$-DCSK SWIPT system has higher spectral efficiency than that of the CI-DCSK SWIPT and SR-DCSK MISO-SWIPT systems.
Furthermore, the spectral efficiency of the proposed CIM-MC-$M$-DCSK system is also equal to $\frac{\log_2N+N\log_2M}{N}$ b/s/Hz. Hence, the proposed system has the same spectral efficiency as the CIM-MC-$M$-DCSK system.



\subsection{Energy Efficiency Analysis}
The energy efficiency $({\rm J/s})$ is defined as the ratio of the spectral efficiency to the total power consumption. The energy efficiency of the proposed CIM-MC-$M$-DCSK MISO-SWIPT system is given by
\begin{flalign}
EE_{P} =  \frac{\log_2N+N\log_2M}{N \left(2NE_1+P_R - P_h \right)}.
\label{eq:EE}
\end{flalign}

Because the CIM-MC-$M$-DCSK SWIPT system has higher spectral efficiency than that of the CI-DCSK SWIPT and SR-DCSK MISO-SWIPT systems, the former system also has higher energy efficiency than that of the latter two systems at the same configures.
Moreover, the energy efficiency of the proposed CIM-MC-$M$-DCSK system is $EE_{Conv} = \frac{\log_2N+N\log_2M}{N\left(2NE_1+P_R\right)}$ J/s. Hence, the proposed system has higher energy efficiency than the CIM-MC-$M$-DCSK system.


\begin{table}[t]
\caption{Configuration of the multipath Rayleigh fading channels.}
\centering\vspace{-1.5mm}
\begin{tabular}{|c|c|c|c|c|c|c|}
\hline
\multirow{2}{*}{Antennas} &
 \multicolumn{3}{c|}{Channel gains} & \multicolumn{3}{c|}{Path delay} \\
 \cline{2-7}
   & $\E[\alpha_{1}^2]$ &  $\E[\alpha_{2}^2]$ & $ \E[\alpha_{3}^2] $ & $\tau_1$ & $\tau_2$ & $\tau_3$\\
 \hline
 $1$ & 0.7 & 0.2 & 0.1 & 0 & 2 & 5\\
  \hline
 $2$ & 0.6 & 0.25 & 0.15 & 0 & 3 & 6 \\
  \hline
 $3$ & 0.8 & 0.12 & 0.08 & 0 & 1 & 2\\
   \hline
 $4$ & 0.28 & 0.42 & 0.3 & 0 & 2 & 4\\
    \hline
\end{tabular}
\end{table}\hspace{-1mm}

\begin{figure}[t]
\centering
\subfigure[\hspace{-0.8cm}]{ \label{fig:subfig:4a} 
\includegraphics[width=3.5in,height=2.6in]{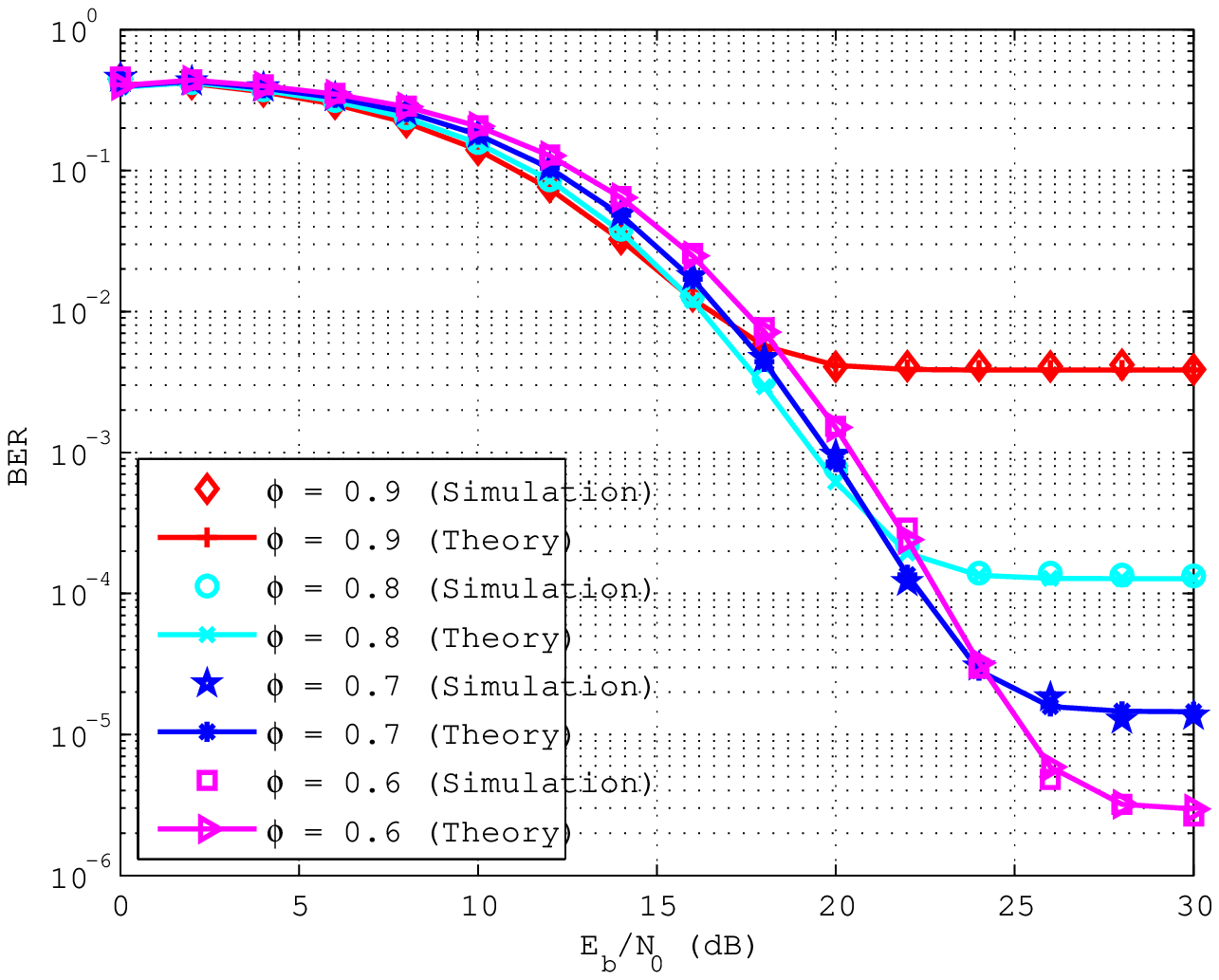}}
\subfigure[\hspace{-0.8cm}]{ \label{fig:subfig:4b} 
\includegraphics[width=3.5in,height=2.6in]{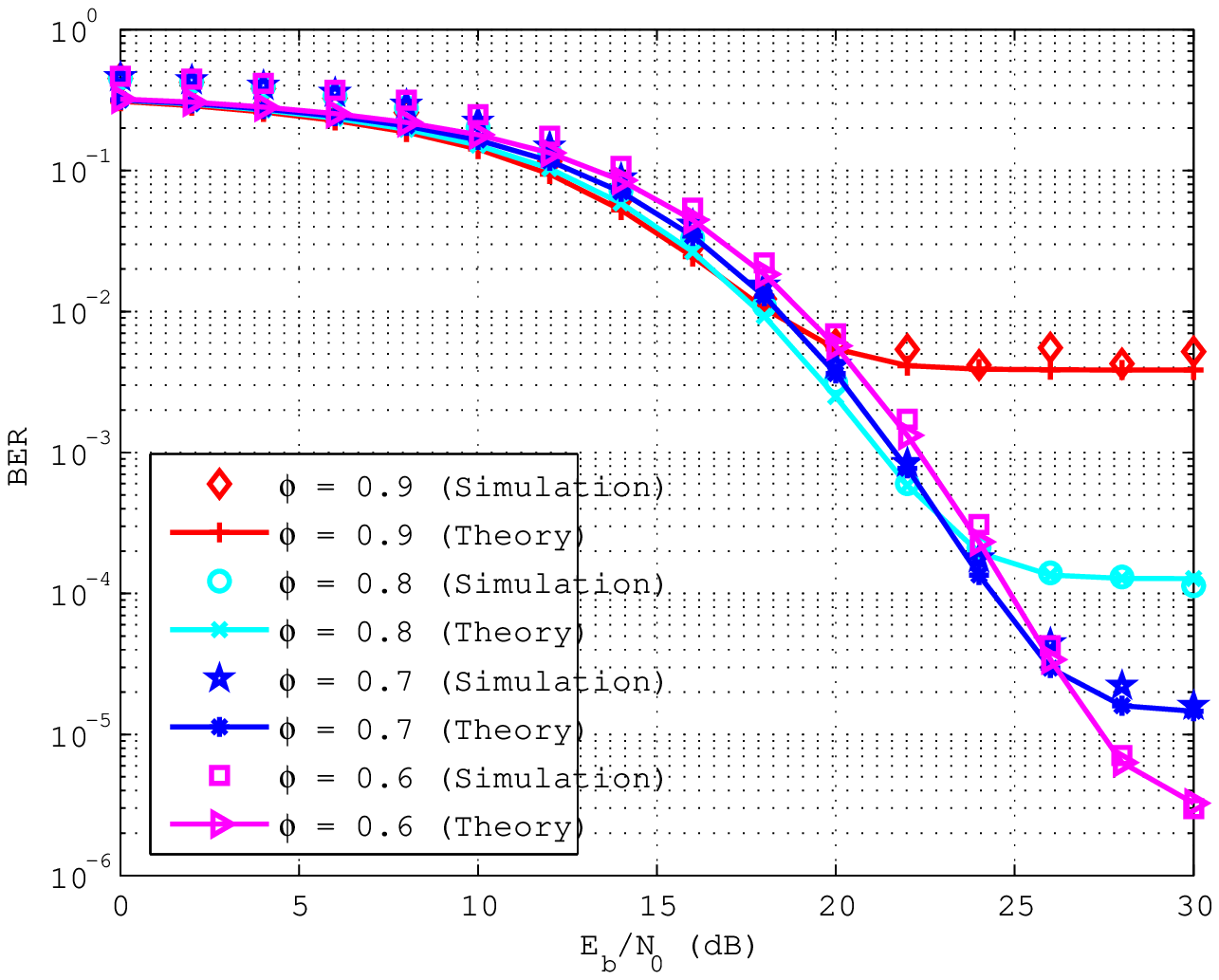}}
\caption{Theoretical and simulated BER results of the MISO-SWIPT CIM-MC-$M$-DCSK system over multipath Rayleigh fading channels, where (a) $N_t = 2$, $M=4$, $N=4$ and (b) $N_t = 2$, $M=8$, $N=8$.}
\label{fig:Fig.4} 
\end{figure}

\begin{figure}[t]
\centering
\subfigure[\hspace{-0.8cm}]{ \label{fig:subfig:5a} 
\includegraphics[width=3.5in,height=2.6in]{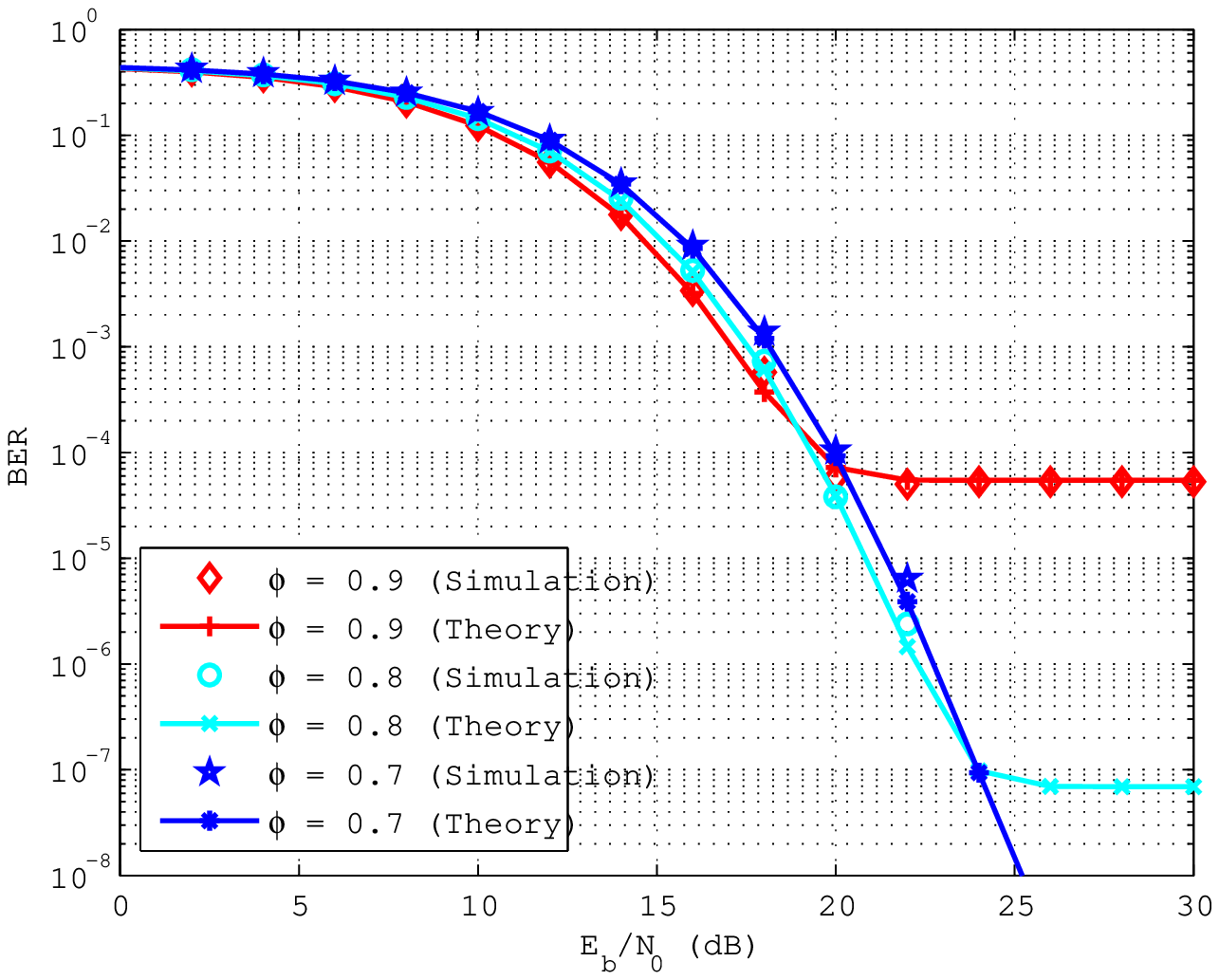}}
\subfigure[\hspace{-0.8cm}]{ \label{fig:subfig:5b} 
\includegraphics[width=3.5in,height=2.6in]{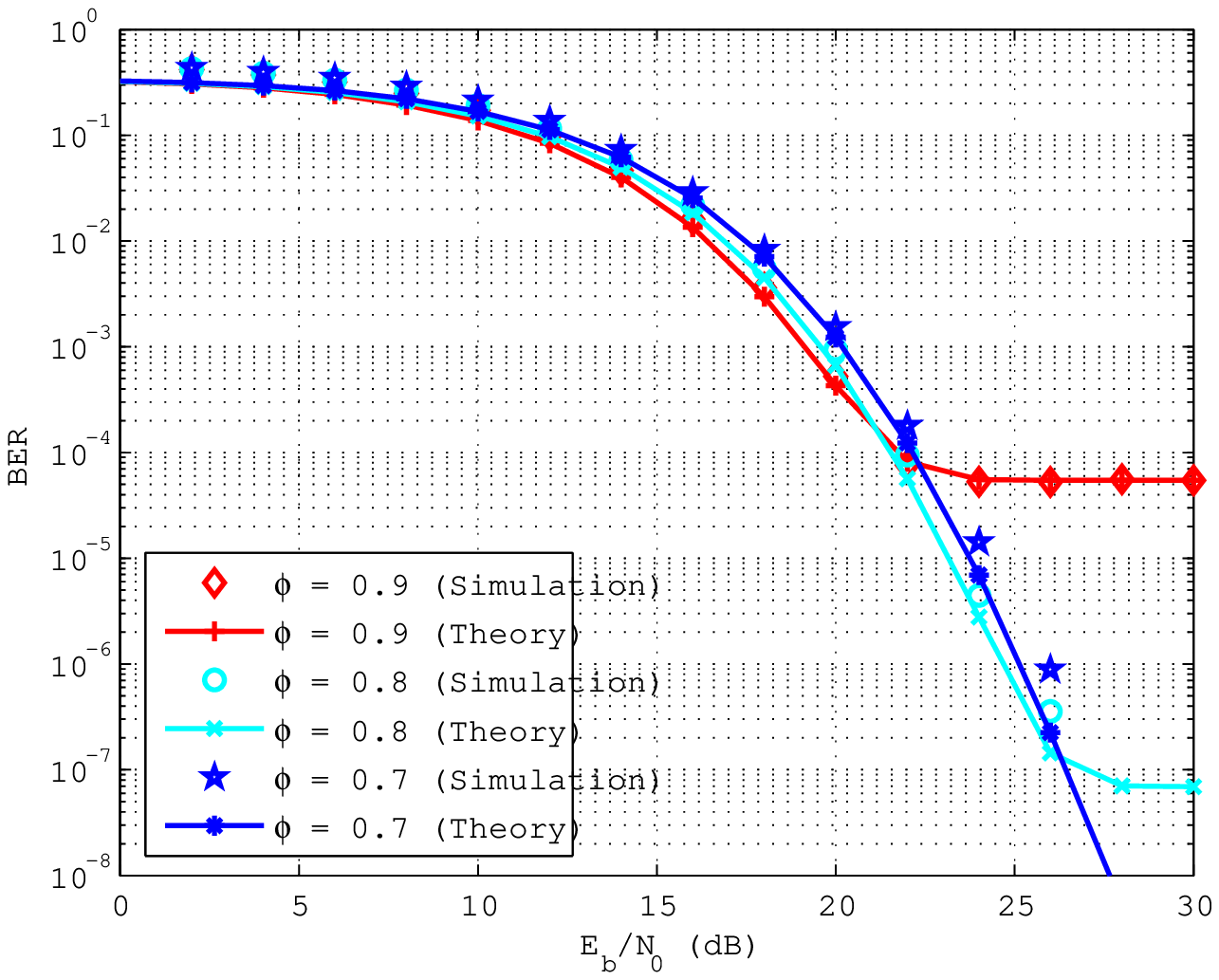}}
\caption{Theoretical and simulated BER results of the MISO-SWIPT CIM-MC-$M$-DCSK system over multipath Rayleigh fading channels, where (a) $N_t = 4$, $M=4$, $N=4$ and (b) $N_t = 4$, $M=8$, $N=8$.}
\label{fig:Fig.5} 
\end{figure}

\begin{figure}[t] 
\center
\includegraphics[width=3.5in,height=2.6in]{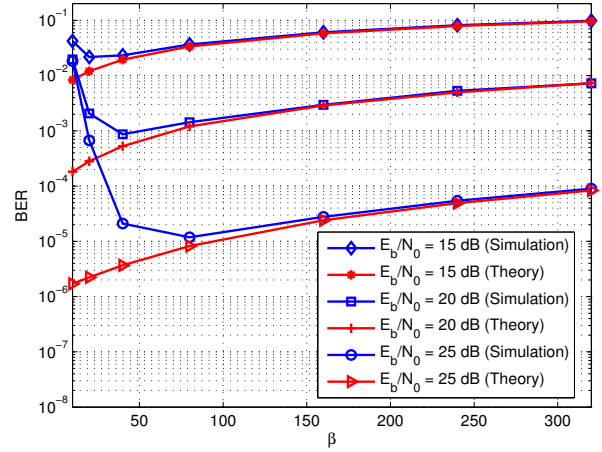}
\caption{The effect of $\beta$ on the BER performance of the proposed system over multipath Rayleigh fading channels.}
\label{fig:Fig.5-1}
\end{figure}

\begin{figure}[t] 
\center
\includegraphics[width=3.5in,height=2.6in]{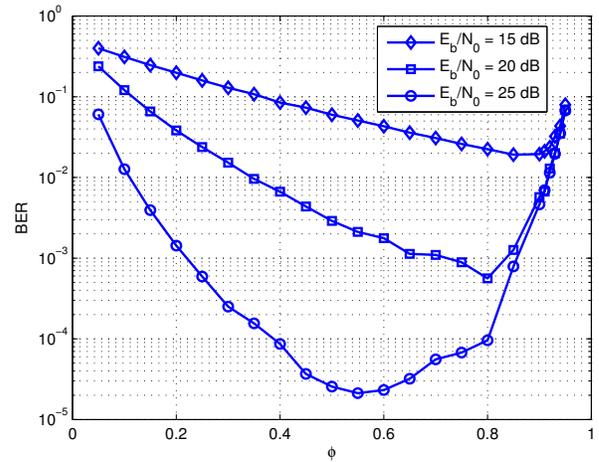}
\caption{BER performance of the proposed system with different value of $\phi$ over multipath Rayleigh fading channels at SNRs of $15$ dB, $20$ dB and $25$ dB.}
\label{fig:Fig.6}
\end{figure}

\begin{figure}[t] 
\center
\includegraphics[width=3.5in,height=2.6in]{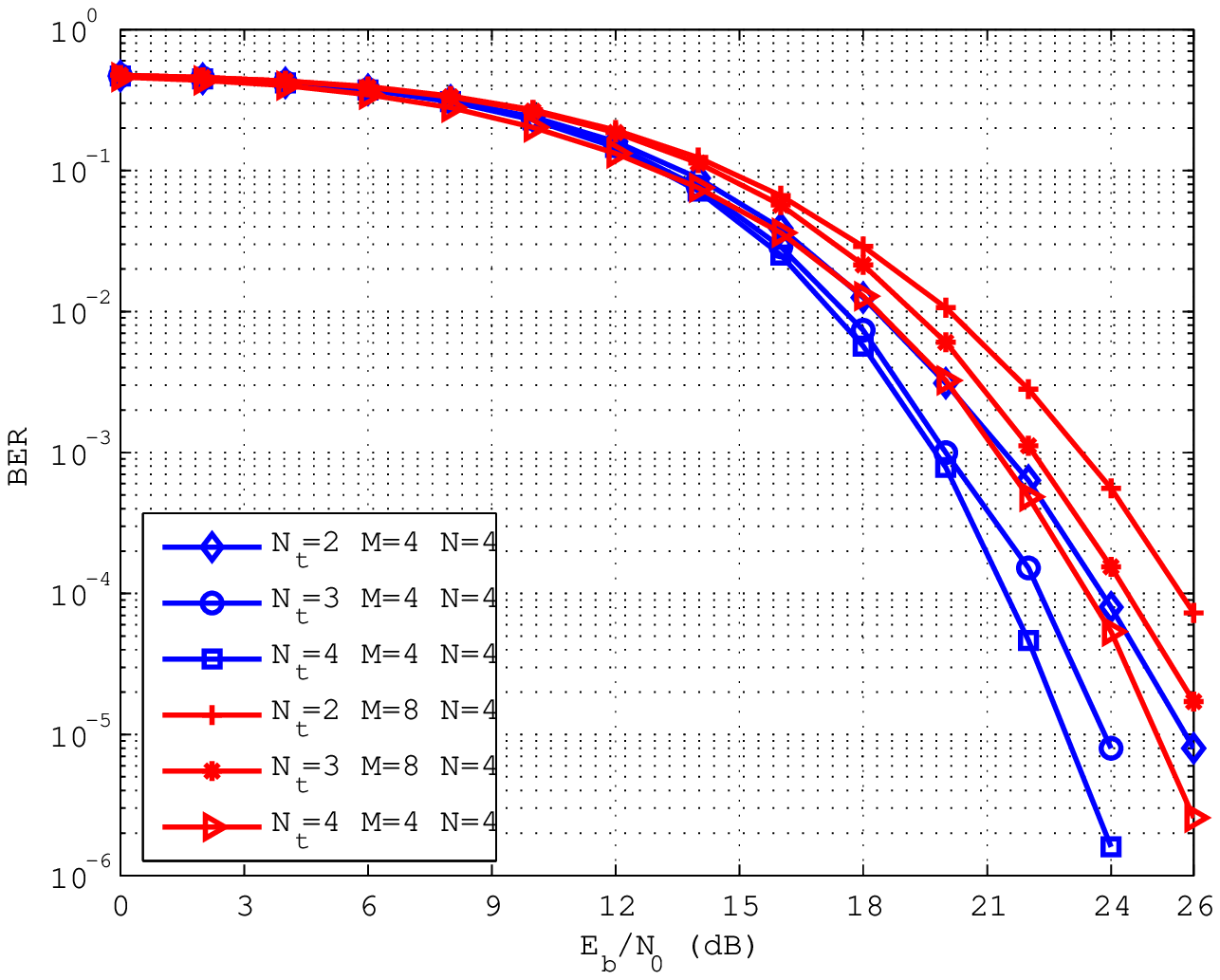}
\caption{BER performance of the proposed system with different value of $N_t$ over multipath Rayleigh fading channels.}
\label{fig:Fig.7}
\end{figure}

\begin{figure}[t] 
\center
\includegraphics[width=3.5in,height=2.6in]{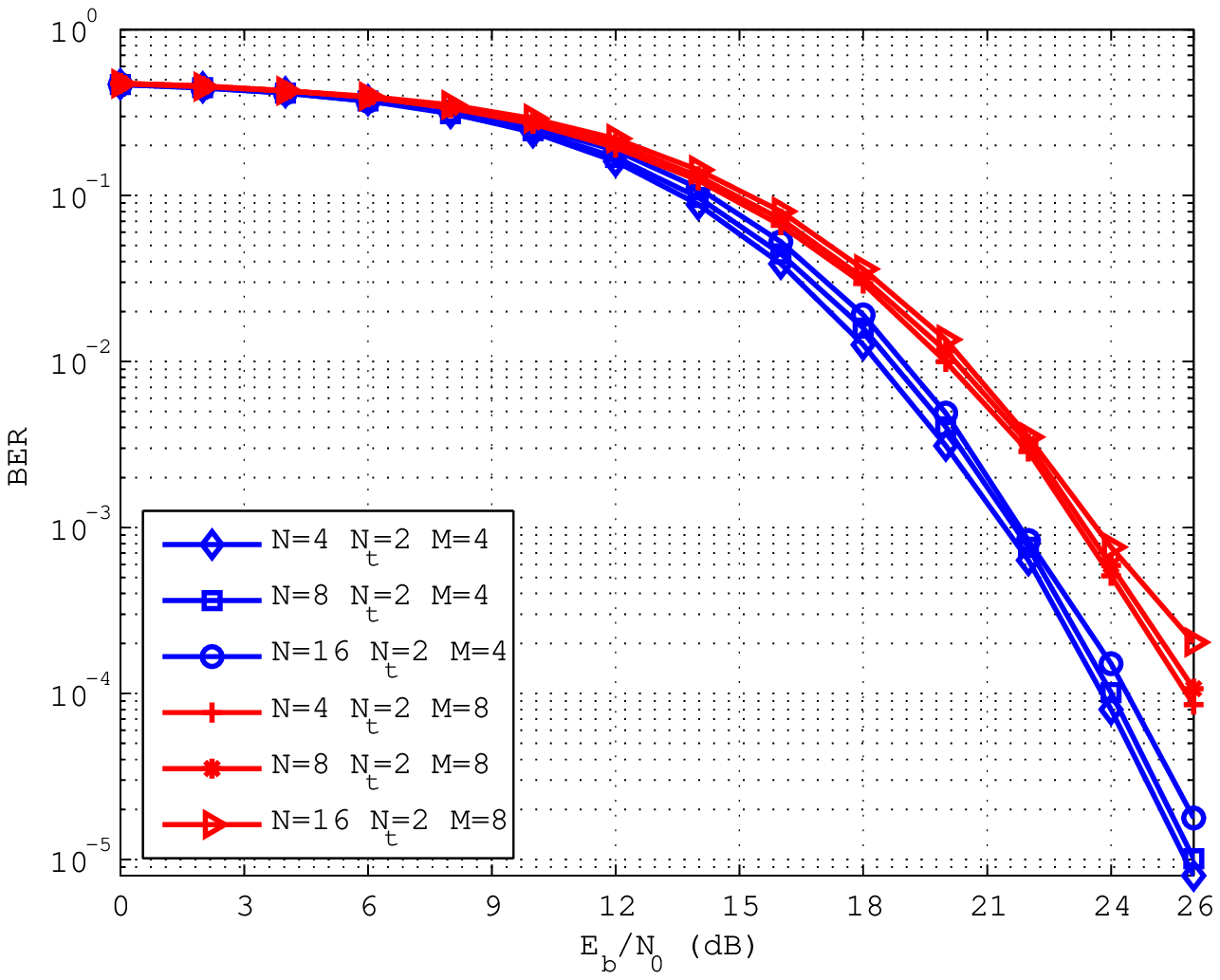}
\caption{BER performance of the proposed system with different value of $N$ over multipath Rayleigh fading channels.}
\label{fig:Fig.8}
\end{figure}

\begin{figure}[t] 
\center
\includegraphics[width=3.5in,height=2.6in]{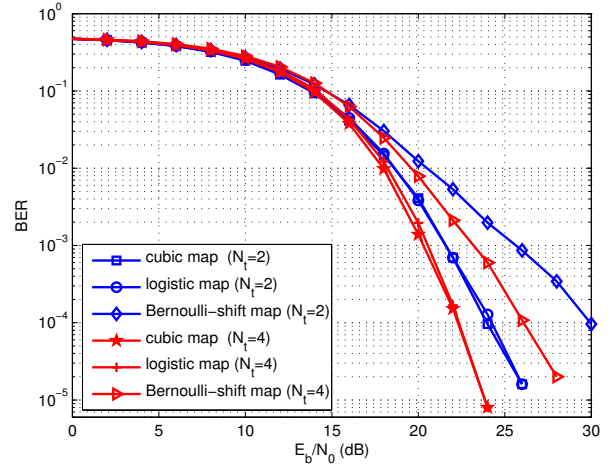}
\caption{The effect of different chaotic signals on the BER performance of the proposed system over multipath Rayleigh fading channels.}
\label{fig:Fig.8-1}
\end{figure}

\begin{figure}[t] 
\center
\includegraphics[width=3.5in,height=2.6in]{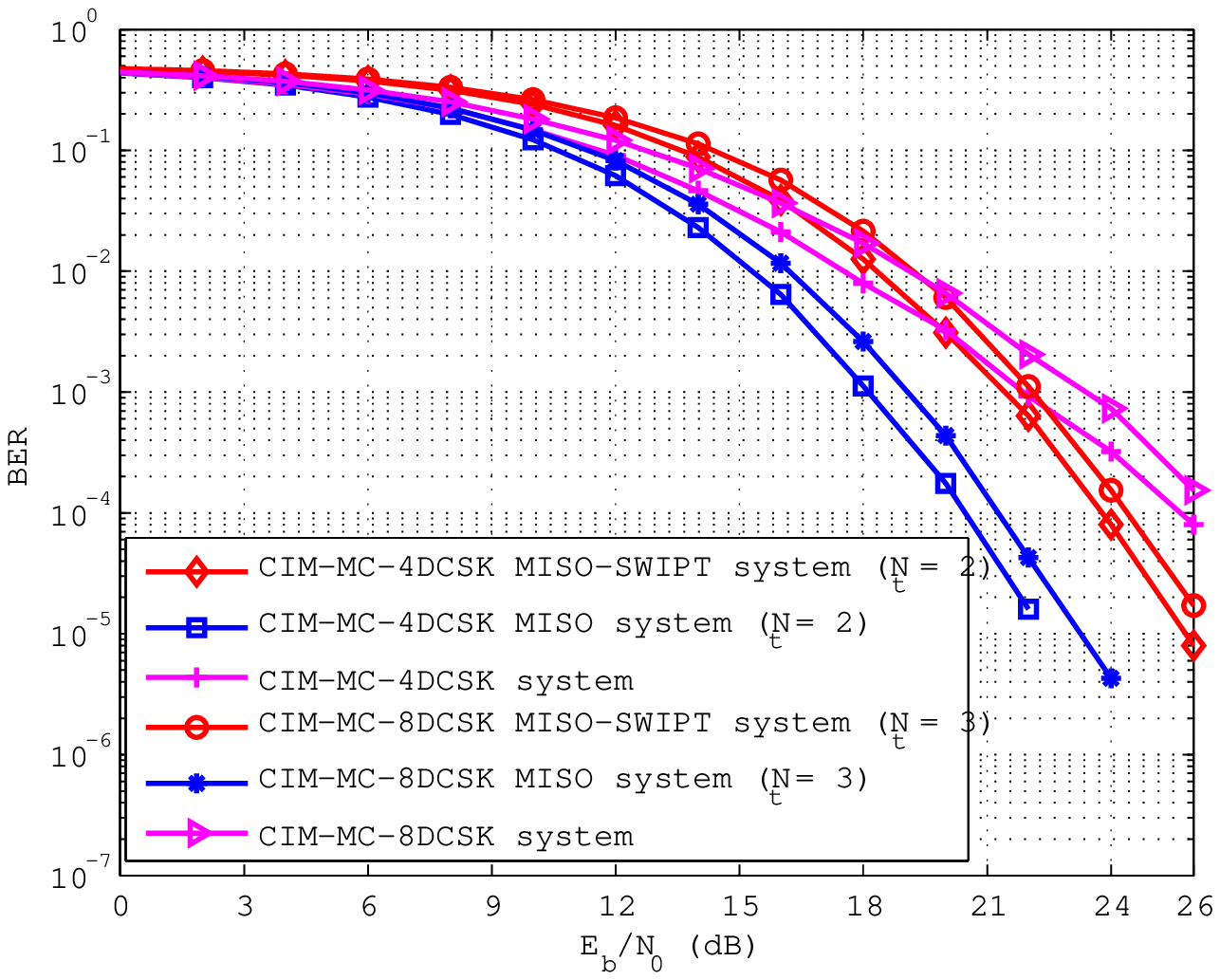}
\caption{BER performance of the proposed system, CIM-MC-$M$-DCSK system and CIM-MC-$M$-DCSK MISO system over multipath Rayleigh fading channels.}
\label{fig:Fig.9}
\end{figure}

\section{Numerical Results and Discussions}\label{sect:res_dis}
We present analytical and simulation plots of the BER for the proposed MISO-SWIPT CIM-MC-$M$-DCSK system  over  multipath Rayleigh fading channels. In simulation, the parameters of the multipath Rayleigh fading channels are given in Table I. Moreover, the spreading factor $\beta$ is set to $160$, the energy-harvesting efficiency $\lambda$ is set to $0.5$ and the power $P_R$ is set to $\frac{2NE_1}{100}$

\subsection{BER Comparison Between Theoretical and Simulated Results}
Figs.~\ref{fig:Fig.4}-\ref{fig:Fig.5} show the theoretical and simulated BER results for the proposed  MISO-SWIPT CIM-MC-$M$-DCSK systems over multipath Rayleigh fading channels, where various number of the antennas, the orders and the subcarriers are adopted. It should be noted that the theoretical results are computed by using Eqs.~\eqref{eq:Eq_7}, \eqref{eq:Eq_27} and \eqref{eq:Eq_38} here.
Referring to the figures, the simulated results are found to agree well with the theoretical results.
In addition, with the decrease of the number of the antennas the proposed system can obtain better system performance due to the diversity.
Moreover, the proposed system has an error floor in terms of BER, which is caused by the energy shortage because the system can not provides enough energy for decoding information of the UE.
Hence, the proposed system can offer a tradeoff between the BER performance and the parameter $\phi$.

Fig.~\ref{fig:Fig.5-1} shows the effect of $\beta$ on the BER performance of the proposed system over multipath Rayleigh fading channels, where $\phi= 0.5$, $N_t=2$, $N=4$ and $M=4$ are adopted.
As can be observed, with the increasing of $\beta$ the theoretical BER performance of the proposed system becomes worse steadily. However, there is an optimal value of $\beta$ to achieve the best simulated BER performance.
The reason is that the ISI is ignored in the BER analysis, while it is not ignored in simulations.
In particular, the ISI becomes smaller as the value of $\beta$ increases. However, with the increasing of $\beta$ it incurs more noise in the system, thus the BER performance deteriorates.

\subsection{Effect of Various Parameters on the System Performance}
The effect of the parameter $\phi$ on the BER performance of the proposed system over multipath Rayleigh fading channels is shown in Fig.~\ref{fig:Fig.6}, where $N_t=2$, $N=4$ and $M=4$ are used. Referring to Fig.~\ref{fig:Fig.6}, there is an optimal value of $\phi$ to achieve best BER performance. This is because, as $\phi$ increases from $0$ to an optimal value, e.g., at a SNR of $20$ dB the optimal value of $\phi$ is $0.8$, most of the energy for the received signal are used for harvesting energy while only a handful of the energy are adopted for decoding information, thus deteriorating the performance of the proposed system. However, as $\phi$ increases from  an optimal value to $1$, only a handful of the energy for the received signal are employed for harvesting energy, thus there is not enough energy for decoding information.  Hence, an adaptive system can be designed to obtain the best performance according to the channel state information.

Fig.~\ref{fig:Fig.7} shows the effect of $N_t$ ($N_t=2,3,4$) on the BER performance of the proposed system over multipath Rayleigh fading channels, where $\phi= 0.5$, $M=4,8$ and $N=4$ are used. It can be observed that the performance of the proposed system with large number of the antennas obtain better performance.
However, as the increase of the number of antennas, the performance improvement slows down.
For example, for $M=4$ and $N=4$, at a BER of $10^{-5}$, the proposed system with three antennas has a $2$ dB gain compared with that with two antennas, while the proposed system with four antennas has a $1$ dB gain compared with that with three antennas. Also, for $M=8$ and $N=4$, at a BER of $10^{-4}$, the proposed system with three antennas has a $1.3$ dB gain compared with that with two antennas, while the proposed system with four antennas has a $0.9$ dB gain compared with that with three antennas. The reason is that, as the increase of the number of the antennas, the ISI is severe due to the imperfect orthogonality of different chaotic signals.

Fig.~\ref{fig:Fig.8} shows the effect of $N$ ($N=4,8,16$) on the BER performance of the proposed system over multipath Rayleigh fading channels, where $\phi= 0.5$, $N_t=2$ and $M=4,8$ are adopted.
Although with increasing the value of $N$ the BER performance of the CIM part improves \cite{46}, referring to Fig.~\ref{fig:Fig.8}, it can be observed that with increasing the value of $N$ BER performance a little deteriorates. This is due to the fact that, as the increase of the value of $N$, the total BER performance of the proposed system is dominated by the BER performance of the $M$-DCSK part. Also, according to Eq.~\eqref{eq:Eq_7}, the above results can be explained. For example, for $N=4$ and $M=4$, $\frac{\log_2N}{\log_2N+N\log_2M}=\frac{1}{5}$ and $\frac{N\log_2M}{\log_2N+N\log_2M}=\frac{4}{5}$, while, for $N=16$ and $M=4$, $\frac{\log_2N}{\log_2N+N\log_2M}=\frac{1}{9}$ and $\frac{N\log_2M}{\log_2N+N\log_2M}=\frac{8}{9}$. Hence, at a large number of subcarriers the total BER performance of the proposed system is determined by the BER performance of the $M$-DCSK part.

We further investigate the effect of different chaotic signals on the BER performance of the proposed system. Logistic map, cubic map and Bernoulli-shift map, which are used as the chaotic signals, are given by \cite{29}, respectively
\begin{align}\nonumber
& \text{logistic map: }{{c }_{k +1}}=1-2c_{k }^{2}, \\\nonumber
& \text{cubic map: }{{c }_{k +1}}=4c _{k }^{3}-3{{c }_{k }}, \\\nonumber
& \text{Bernoulli-shift map: }{{c }_{k +1}}=\left\{ \begin{aligned}
& 1.2{{c}_{k }}+1,\text{when }{{c }_{k }}<0 \\\nonumber
& 1.2{{c }_{k }}-1,\text{when }{{c }_{k }}>0 \nonumber
\end{aligned} \right..\nonumber
\end{align}
Fig.~\ref{fig:Fig.8-1} shows the effect of different chaotic signals on the BER performance of the proposed system, where $\phi= 0.5$, $N=4$, $M=4$, $N_t=2$ and $N_t=4$ are adopted.
Referring to this figure, the proposed systems with the logistic and cubic maps have the same BER performance while the one with Bernoulli-shift map has worse BER performance. In particular, it can be seen that the proposed systems with the logistic and cubic maps have a $6$-dB gain compared with the one with Bernoulli-shift map. Hence, it is very important to suitably select a chaotic map to generate chaotic signal in practical applications.

\subsection{Comparison between the Proposed System and Existing Counterparts}

As a further insight, the BER performance of the proposed system is compared to the CIM-MC-$M$-DCSK system and CIM-MC-$M$-DCSK MISO system (i.e., no energy harvesting), as shown in Fig.~\ref{fig:Fig.9}, where $N=4$. Referring to this figure, although the proposed system has a BER performance loss compared with the CIM-MC-$M$-DCSK MISO system at the same spectral efficiency, e.g., about $2.5$ dB performance loss at a BER of $4 \times 10^{-5}$, the proposed system can offer self-sustainable ability  for power supply, i.e., it does not require any external power supply.
Furthermore, the proposed system can offer better BER performance compare with the CIM-MC-$M$-DCSK system at the high SNR regime.
For example, at a BER of $10^{-4}$, the CIM-MC-8DCSK MISO-SWIPT system can achieves a $2$~dB gain compared with the the CIM-MC-8DCSK system.






\section{Conclusions}\label{sect:conclusion}
In this paper, we have proposed a new MISO-SWIPT scheme for the CIM-MC-$M$-DCSK system, which can simultaneously provide energy and transmit information for the UEs without any external
power supply.
The proposed system not only inherit the low-complexity and strong anti-multipath-fading capability advantages of the CIM-MC-$M$-DCSK system, but also achieves sustainable power supply for the medical devices.
Moreover, the performance of the proposed system has been carefully analyzed in terms of BER, spectral efficiency and energy efficiency.
Through analysis and simulations, the following results have been obtained: 1)~the proposed system exhibits higher spectral efficiency compared with the SR-DCSK SWIPT and CI-DCSK SWIPT systems, while it offers the same spectral efficiency compared to the CIM-MC-$M$-DCSK system; 2)~the proposed system achieves higher energy efficiency than the the SR-DCSK SWIPT, CI-DCSK SWIPT and CIM-MC-$M$-DCSK system; 3)~the proposed system possesses self-sustainable ability and better BER performance compared with the CIM-MC-$M$-DCSK system.
Thanks to the above benefits, the proposed CIM-MC-$M$-DCSK MISO-SWIPT system can be considered as an excellent candidate for the low-cost, battery-capacity-limited and low-power e-health IoT applications.
In the future, we will further investigate the multi-user and uplink scenarios for the proposed CIM-MC-$M$-DCSK MISO-SWIPT system.

\appendices
\section{Means and variances of \eqref{eq:Eq_13} and \eqref{eq:Eq_17}}
The means and variances of \eqref{eq:Eq_13} and \eqref{eq:Eq_17} are derived as follows.

Eq.~\eqref{eq:Eq_13} is rewritten as
\begin{align}
z_{e,j} &=  \sum_{k=1}^{\beta} \Big{(} \sqrt{\phi} N X_k  + Y_k \Big{)}^2 \nonumber \\
&= \sum_{k=1}^{\beta} \Big{(} \phi^2N^2 X_{k}^2 + 2 \sqrt{\phi} N X_k Y_k + Y_{k}^2 \Big{)}
\label{eq:Eq_A_1}
\end{align}
where
\vspace{-3.5mm}
\begin{align}
X_k = \sum_{n_t=1}^{N_t} \sum_{l=1}^{L_{n_t}} \frac{\alpha_{l,n_t} }{\sqrt{N_t}}  c_{x,k-\tau_{l,n_t}}^{n_t},
\label{eq:Eq_A_2}
\end{align}
\vspace{-3.5mm}
\begin{align}
Y_k = \sum_{i=1}^{N}w_{j,i}n_{x,k}^{i}.
\label{eq:Eq_A_3}
\end{align}

The mean and variance of Eq.~\eqref{eq:Eq_A_1} is given by
\begin{align}
\E \left[ z_{e,j} \right] =  & \phi^2N^2 \E \left[ \sum_{k=1}^{\beta} X_{k}^2 \right] + 2 \sqrt{\phi} N \E \left[ \sum_{k=1}^{\beta} X_k Y_k \right]  \nonumber \\
&+  \E \left[ \sum_{k=1}^{\beta} Y_{k}^2 \right],
\label{eq:Eq_A_1_1}
\end{align}
\vspace{-5.5mm}
\begin{align}
Var \left[ z_{e,j} \right] =  & \phi^4N^4 Var \left[ \sum_{k=1}^{\beta} X_{k}^2 \right] \hspace{-1.5mm} + 4 \phi N^2 Var \left[ \sum_{k=1}^{\beta} X_k Y_k \right]  \nonumber \\
&+  Var \left[ \sum_{k=1}^{\beta} Y_{k}^2 \right].
\label{eq:Eq_A_1_2}
\end{align}

Because of the orthogonality of different chaotic signals, one has the following approximated expression:
\begin{align}
\sum_{k=1}^{\beta} \left( \sum_{l=1}^{L_{i}} \alpha_{l,i} c_{x,k-\tau_{l,i}}^{i} \times \sum_{l=1}^{L_{j}} \alpha_{l,j}  c_{x,k-\tau_{l,j}}^{i} \right) \approx 0, i\ne j,
\label{eq:Eq_A_4}
\end{align}
Additionally, according to \cite{25,26}, for a large spreading factor, one also has the following approximated expression:
\vspace{-1.5mm}
\begin{align}
\sum_{k=1}^{\beta} c_{x,k-\tau_{i,n_t}}^{n_t} c_{x,k-\tau_{j,n_t}}^{n_t} \approx 0, i \ne j.
\label{eq:Eq_A_5}
\end{align}

Using Eqs.~\eqref{eq:Eq_A_4} and \eqref{eq:Eq_A_5}, one can obtain as
\begin{align}
\sum_{k=1}^{\beta} X_{k}^2 &= \sum_{k=1}^{\beta} \left(  \sum_{n_t=1}^{N_t} \sum_{l=1}^{L_{n_t}} \frac{\alpha_{l,n_t} }{\sqrt{N_t}}  c_{x,k-\tau_{l,n_t}}^{n_t} \right)^2 \nonumber \\
& \approx  \sum_{k=1}^{\beta} \sum_{n_t=1}^{N_t} \left( \sum_{l=1}^{L_{n_t}} \frac{\alpha_{l,n_t} }{\sqrt{N_t}}  c_{x,k-\tau_{l,n_t}}^{n_t} \right)^2 \nonumber \\
& \approx  \sum_{k=1}^{\beta} \sum_{n_t=1}^{N_t} \sum_{l=1}^{L_{n_t}} \frac{\alpha_{l,n_t}^2 }{N_t}  \left( c_{x,k-\tau_{l,n_t}}^{n_t} \right)^2 \nonumber \\
& = \frac{E_1 }{N_t}  \sum_{n_t=1}^{N_t} \sum_{l=1}^{L_{n_t}} \alpha_{l,n_t}^2.
\label{eq:Eq_A_6}
\end{align}

The mean and variance of Eq.~\eqref{eq:Eq_A_6} are obtained as
\begin{align}
\E \left[ \sum_{k=1}^{\beta} X_{k}^2 \right] = \frac{E_1 }{N_t}  \sum_{n_t=1}^{N_t} \sum_{l=1}^{L_{n_t}} \alpha_{l,n_t}^2, Var\left[ \sum_{k=1}^{\beta} X_{k}^2 \right] = 0.
\label{eq:Eq_A_7}
\end{align}

Furthermore, one has
\begin{align}
\sum_{k=1}^{\beta} X_{k}Y_k  &= \sum_{k=1}^{\beta} \left( \sum_{n_t=1}^{N_t} \sum_{l=1}^{L_{n_t}} \frac{\alpha_{l,n_t} }{\sqrt{N_t}}  c_{x,k-\tau_{l,n_t}}^{n_t} \hspace{-1.5mm} \times \sum_{i=1}^{N} w_{j,i}n_{x,k}^{i}\right) \nonumber \\
&=\sum_{k=1}^{\beta} \sum_{i=1}^{N} \left( \sum_{n_t=1}^{N_t} \sum_{l=1}^{L_{n_t}} \frac{\alpha_{l,n_t} }{\sqrt{N_t}}  c_{x,k-\tau_{l,n_t}}^{n_t} w_{j,i}n_{x,k}^{i}\right).
\label{eq:Eq_A_8}
\end{align}
The mean and variance of Eq.~\eqref{eq:Eq_A_8} can be calculated as
\begin{align}
&\E \left[ \sum_{k=1}^{\beta} X_{k}Y_k \right] =0, \nonumber \\
&Var \left[ \sum_{k=1}^{\beta} X_{k}Y_k \right] = \sum_{k=1}^{\beta} \sum_{i=1}^{N} \E \left[ X_{k}^2\right] \E \left[ \left( n_{x,k}^{i} \right)^2 \right] \nonumber \\
&~~~~~~~~~~ ~~~~~~~~~~~~ =\frac{NN_0E_1}{2N_t} \sum_{n_t=1}^{N_t} \sum_{l=1}^{L_{n_t}} \alpha_{l,n_t}^2.
\label{eq:Eq_A_9}
\end{align}

Moreover, because $n_{x,k}^i$ follows Gaussian distribution with zero mean and the variance $\frac{N_0}{2}$, the variable $Y_k$ also follows the Gaussian distribution with zero mean and the variance $\frac{NN_0}{2}$.
Hence, the variable $\sum_{k=1}^{\beta} Y_{k}^2$ is a Chi-square distribution with $\beta$ degrees of freedom, and its mean and variance is given by
\begin{align}
\E \left[ \sum_{k=1}^{\beta} Y_{k}^2 \right] = \frac{\beta NN_0}{2}, Var \left[ \sum_{k=1}^{\beta} Y_{k}^2 \right] =  \frac{\beta N^2 N_{0}^2}{2}.
\label{eq:Eq_A_10}
\end{align}

As a consequence, substituting Eqs.~\eqref{eq:Eq_A_7}, \eqref{eq:Eq_A_9} and \eqref{eq:Eq_A_10} into \eqref{eq:Eq_A_1_1} and \eqref{eq:Eq_A_1_2} yields the expressions \eqref{eq:Eq_14} and \eqref{eq:Eq_15}.
Similarly, the mean and variance of Eq.~\eqref{eq:Eq_17} can be computed.

\section{Means and variances of \eqref{eq:Eq_28} and \eqref{eq:Eq_29}}
The means and variances of \eqref{eq:Eq_28} and \eqref{eq:Eq_29} are derived as follows.

Eq.~\eqref{eq:Eq_28} is rewritten as,
\begin{align}
\hspace{-1.5mm} z_{a,\hat{S}_i} & \approx  \sum_{k=1}^{\beta} \Big{(} \sqrt{\phi} X_k \hspace{-1mm} + \hspace{-1mm} n_{x,k}^{1} \Big{)} \hspace{-1.5mm} \times \hspace{-1.5mm} \Big{(} \sqrt{\phi} a_{S_i} X_k \hspace{-1mm} + \hspace{-1mm} \sqrt{\phi} b_{S_i} Z_k \hspace{-1mm} + \hspace{-1mm} n_{x,k+\beta}^{i} \Big{)} \nonumber \\
&=  \phi a_{S_i} \sum_{k=1}^{\beta} X_{k}^2 \hspace{-1mm} + \hspace{-1mm} \phi b_{S_i} \sum_{k=1}^{\beta} X_{k}Z_k \hspace{-1mm} + \hspace{-1mm} \sqrt{\phi} \sum_{k=1}^{\beta} X_{k} n_{x,k+\beta}^{i} + \nonumber\\
&~~~ \hspace{-1.5mm} \sqrt{\phi} a_{S_i} \sum_{k=1}^{\beta} X_{k} n_{x,k}^{1} \hspace{-1mm} + \hspace{-1mm} \sqrt{\phi} b_{S_i} \sum_{k=1}^{\beta} Z_{k} n_{x,k}^{1} \hspace{-1mm} + \hspace{-1mm} \sum_{k=1}^{\beta} n_{x,k}^{1} n_{x,k+\beta}^{i},
\label{eq:Eq_B_1}
\end{align}
where
\begin{align}
Z_k = \sum_{n_t=1}^{N_t} \sum_{l=1}^{L_{n_t}} \frac{\alpha_{l,n_t} }{\sqrt{N_t}}  c_{y,k-\tau_{l,n_t}}^{n_t}.
\label{eq:Eq_B_2}
\end{align}

According to Eqs.~\eqref{eq:Eq_A_4} and \eqref{eq:Eq_A_5}, one has
\begin{align}
\phi b_{S_i} \sum_{k=1}^{\beta} X_{k}Z_k \approx 0.
\label{eq:Eq_B_3}
\end{align}

The mean of Eq.~\eqref{eq:Eq_B_1} can be calculated as
\begin{align}
\E \left[ z_{a,\hat{S}_i} \right] &=  \E \left[ \phi a_{S_i} \sum_{k=1}^{\beta} X_{k}^2 \right] = \frac{\phi a_{S_i}E_1 }{N_t}  \sum_{n_t=1}^{N_t} \sum_{l=1}^{L_{n_t}} \alpha_{l,n_t}^2,
\label{eq:Eq_B_4}
\end{align}

The variance of Eq.~\eqref{eq:Eq_B_1} is computed as
\begin{align}
\hspace{-1.5mm} &Var \hspace{-1mm}\left[ z_{a,\hat{S}_i} \hspace{-1mm} \right] \hspace{-1mm} =
\phi Var \hspace{-1mm}\left[ \sum_{k=1}^{\beta} X_{k} n_{x,k+\beta}^{i} \hspace{-1mm} \right] \hspace{-1mm} + \hspace{-1mm} \phi a_{S_i}^2 Var \left[ \sum_{k=1}^{\beta} X_{k} n_{x,k}^{1} \right] \nonumber\\
& + \phi b_{S_i}^2 Var \left[ \sum_{k=1}^{\beta} Z_{k} n_{x,k}^{1} \right] + Var \left[ \sum_{k=1}^{\beta} n_{x,k}^{1} n_{x,k+\beta}^{i} \right] \nonumber \\
\hspace{-1.5mm}& = \phi \sum_{k=1}^{\beta} \E  \left[ X_{k}^2 \right] \E \hspace{-0.5mm}\left[ \left( n_{x,k+\beta}^{i} \right)^2 \hspace{-0.5mm} \right] \hspace{-1mm} + \hspace{-1mm}
\phi a_{S_i}^2 \sum_{k=1}^{\beta} \E \hspace{-0.5mm} \left[ X_{k}^2 \right] \E \left[ \left(n_{x,k}^{1} \right)^2 \hspace{-0.5mm} \right] \nonumber \\
& \hspace{-1mm} + \hspace{-1mm} \phi b_{S_i}^2  \sum_{k=1}^{\beta} \E \left[ Z_{k}^2 \right] \E \hspace{-0.5mm} \left[ \left( n_{x,k}^{1}\right)^2 \hspace{-0.5mm}\right]\hspace{-1mm} + \hspace{-1mm} \sum_{k=1}^{\beta} \E \hspace{-0.5mm} \left[ \left( n_{x,k}^{1} \right)^2 \hspace{-0.5mm} \right] \E \hspace{-0.5mm} \left[ \left( n_{x,k+\beta}^{i} \right)^2 \hspace{-0.5mm} \right] \nonumber \\
& = \frac{\phi N_0 E_1}{N_t} \sum_{n_t=1}^{N_t} \sum_{l=1}^{L_{n_t}} \alpha_{l,n_t}^2 + \frac{\beta N_{0}^2}{4}.
\label{eq:Eq_B_5}
\end{align}

Therefore, the mean and variance of \eqref{eq:Eq_28} are obtained in Eq.~\eqref{eq:Eq_B_4} and Eq.~\eqref{eq:Eq_B_5}, respectively. Similarly, the mean and variance of Eq.~\eqref{eq:Eq_29} can be calculated.


\begin{IEEEbiography}[{\includegraphics[width=1.1in,height=1.25in,clip,keepaspectratio]{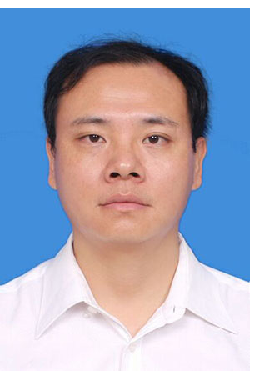}}]
{Guofa Cai} (M'17) received the B.S. degree in communication engineering from Jimei University, Xiamen, China, in 2007, the M.S. degree in circuits and systems from Fuzhou University, Fuzhou, China, in 2012, and the Ph.D. degree in communication engineering from Xiamen University, Xiamen, China, in 2015. In 2017, he was a Research Fellow at the School of Electrical and Electronic Engineering, Nanyang Technological University, Singapore. He is currently an Associate Professor with the School of Information Engineering, Guangdong University of Technology, China. His primary research interests include information theory and coding, spread-spectrum modulation, wireless body area networks, and Internet of Things.
\end{IEEEbiography}

\begin{IEEEbiography}[{\includegraphics[width=1in,height=1.25in,clip,keepaspectratio]{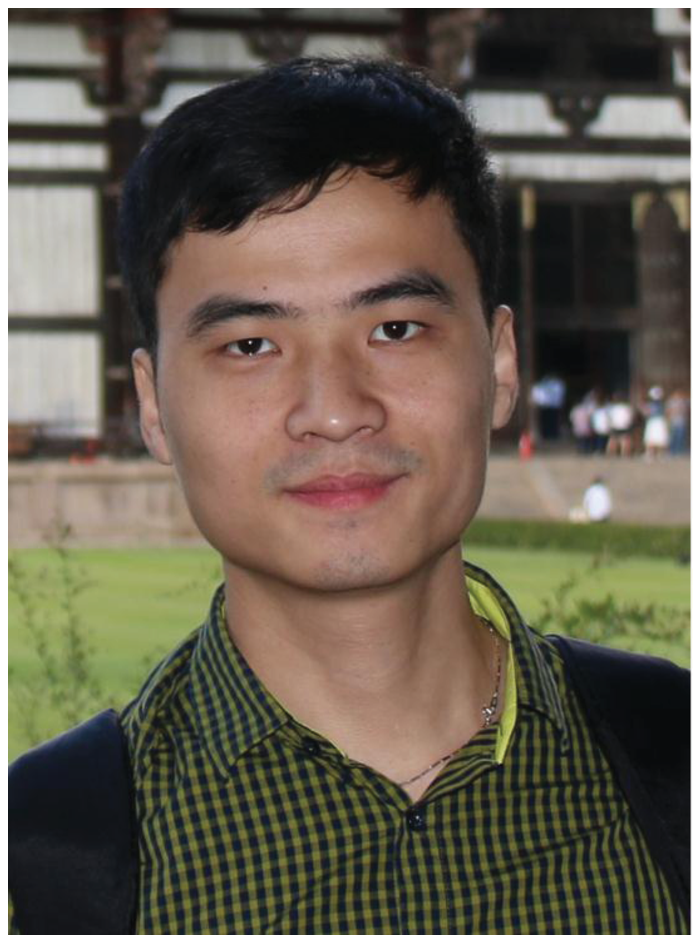}}]
{Yi Fang} (M'15)  received the B.Sc. degree in electronic engineering from East China Jiaotong University, China, in 2008, and the Ph.D. degree in communication engineering,
Xiamen University, China, in 2013. From May 2012 to July 2012, He was a Research Assistant in electronic and information engineering, Hong Kong Polytechnic University, Hong Kong. From September 2012 to September 2013, he was a Visiting Scholar in electronic and electrical engineering, University College London, UK. From February 2014 to February 2015, he was a Research Fellow at the School of Electrical and Electronic Engineering, Nanyang Technological University, Singapore. He is currently a Full Professor at the School of Information Engineering, Guangdong University of Technology, China. He has been an Associate Editor for the IEEE Access since 2018 and a core member of the Guangdong Innovative Research Team since 2016. He served as the Publicity Co-Chair of the International Symposium on Turbo Codes and Iterative Information Processing 2018. His current research interests include information and coding theory (especially LDPC codes), spread-spectrum modulation, and cooperative communications. He is the corresponding author of this article.
\end{IEEEbiography}

\begin{IEEEbiography}
[{\includegraphics[width=1.1in,height=1.25in,clip,keepaspectratio]{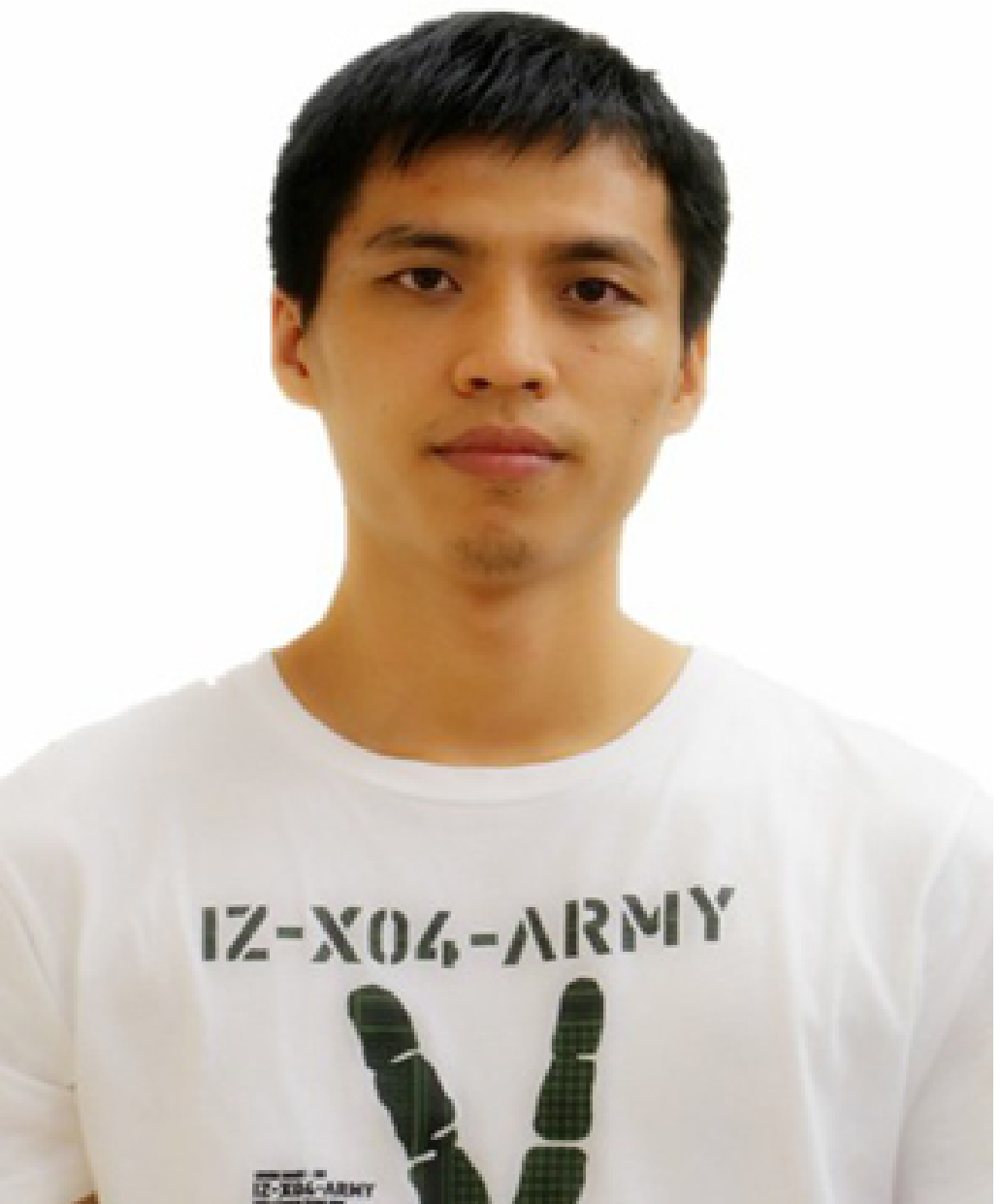}}]
{Pingping Chen} received the Ph.D. degree in electronic engineering from Xiamen University, China, in 2013. In 2012, he was a Research Assistant in electronic and information engineering with The Hong Kong Polytechnic University, Hong Kong. From 2013 to 2015, he was a Post-Doctoral Fellow at the Institute of Network Coding, The Chinese University of Hong Kong, Hong Kong. He is currently a Professor with Fuzhou University, China. His primary research interests include channel coding, joint source and channel coding, network coding, and UWB communications.
\end{IEEEbiography}

\begin{IEEEbiography}[{\includegraphics[width=1.1in,height=1.25in,clip,keepaspectratio]{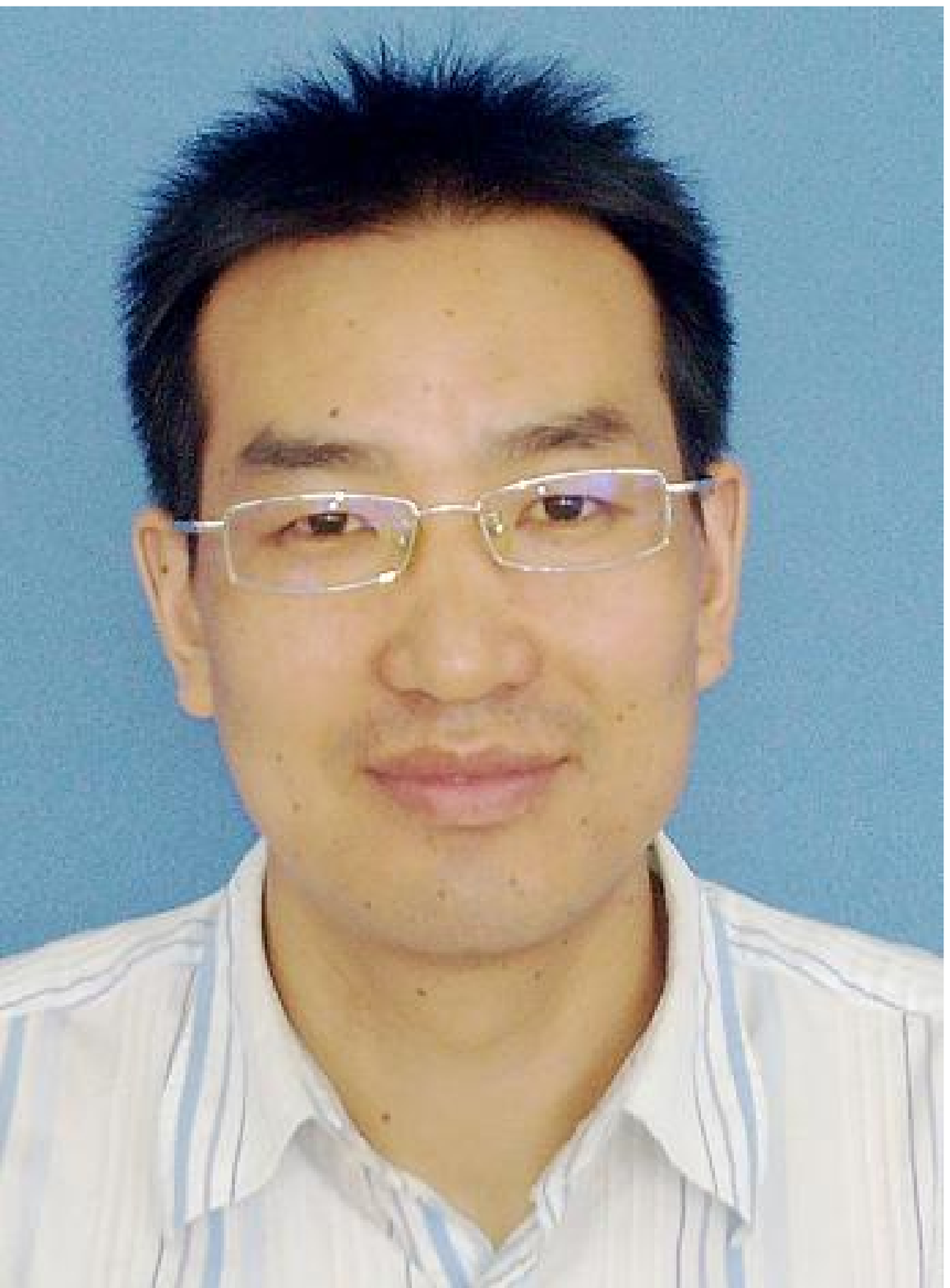}}]
{Guojun Han} (M'12-SM'14) obtained his Ph.D. from Sun Yat-sen University, Guangzhou, China, and the M.E. degree in electronic engineering from South China University of Technology, Guangzhou, China. From March 2011 to August 2013, he was a Research Fellow at the School of Electrical and Electronic Engineering, Nanyang Technological University, Singapore. From October 2013 to April 2014, he was a Research Associate at the Department of
Electrical and Electronic Engineering, Hong Kong University of Science and Technology. He is now a Full Professor and Executive Dean at the School of Information Engineering, Guangdong University of Technology, Guangzhou, China. His research interests include wireless communications, coding and signal processing for data
storage.
\end{IEEEbiography}

\begin{IEEEbiography}[{\includegraphics[width=1.1in,height=1.25in,clip,keepaspectratio]{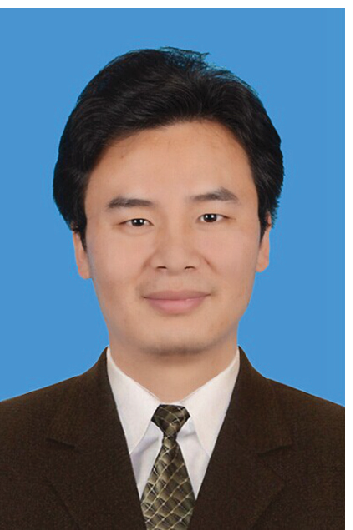}}]
{Guoen Cai}  received the B.Sc in medicine from Fujian Medical University in 2005 and the M.Sc. degree in medicine from the Shanghai Jiao Tong University School of Medicine in 2008. He is currently an Attending Physician with the Fujian Medical University Union Hospital, Fuzhou, China. His primary research interests include Parkinson's disease, virtual reality in medicine, wearable devices in medicine and wireless body area networks.
\end{IEEEbiography}

\begin{IEEEbiography}[{\includegraphics[width=1.1in,height=1.25in,clip,keepaspectratio]{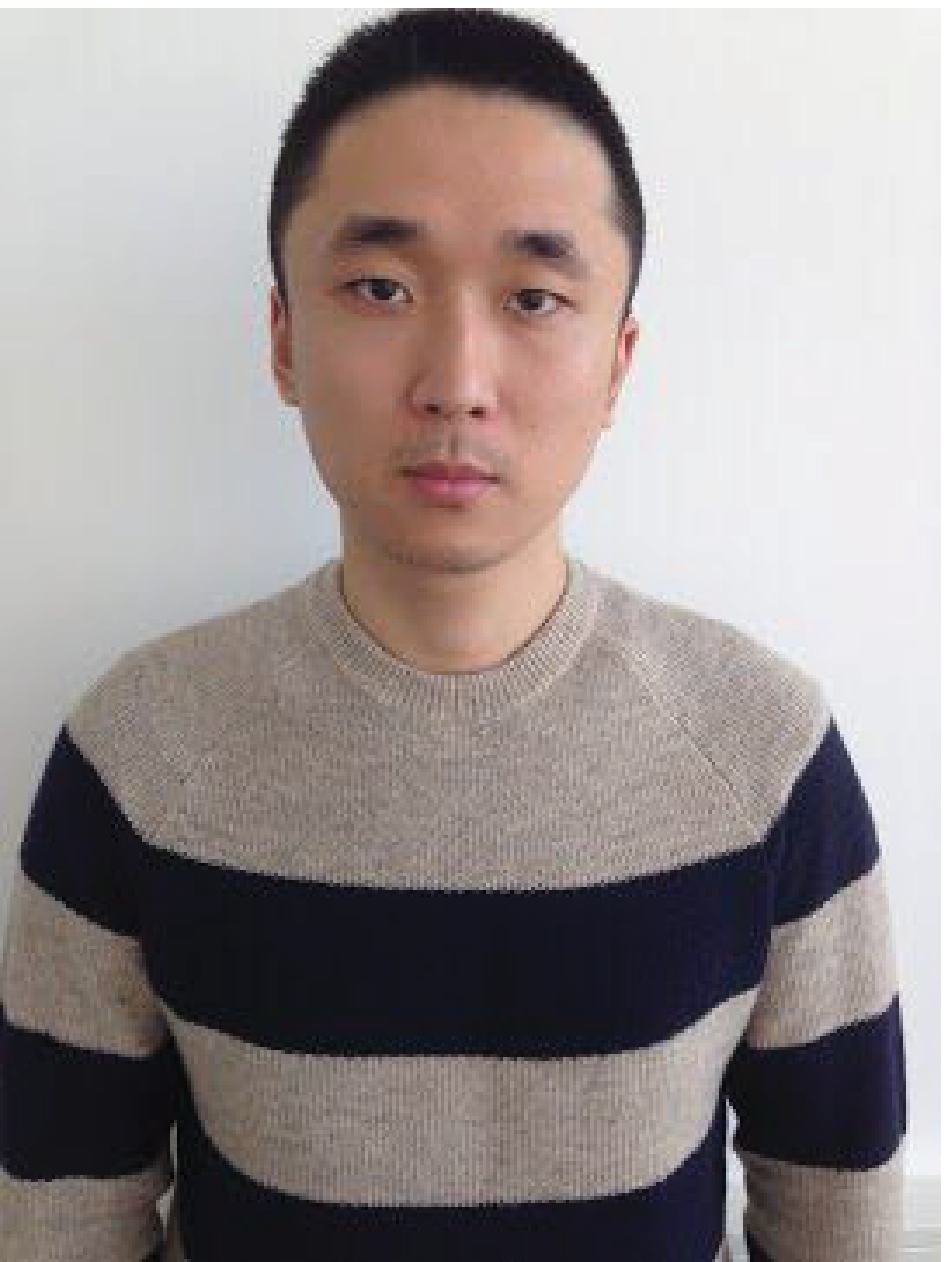}}]
{Yang Song} received the B.Eng. degree in communication engineering from Zhejiang University City College, Hangzhou, China, in 2007, and the M.Eng. and Ph.D. degrees in electronic and information engineering from The Hong Kong Polytechnic University, Hong Kong, in 2008 and 2013, respectively. He was a Research Associate with The Hong Kong Polytechnic University until 2014. From 2014 to 2016, he was a Post-Doctoral Research Associate with the Universit$\ddot{a}$t Paderborn, Paderborn, Germany. Since 2016, he has been a Research Fellow with Nanyang Technological University, Singapore. His current research interests include space-time signal processing. He is an Associate Editor of IET Signal Processing.
\end{IEEEbiography}

\end{document}